\documentclass[a4paper,fleqn,usenatbib]{mnras}

\usepackage{newtxtext,newtxmath}

\usepackage[T1]{fontenc}
\usepackage{ae,aecompl}

\usepackage{graphicx}	
\usepackage{amsmath}	
\usepackage{amssymb}	
\usepackage{multirow}
\usepackage{booktabs}
\usepackage{threeparttable}

\newcommand{\kms}{km~s$^{-1}$}
\newcommand{\cmN}{cm$^{-2}$}
\newcommand{\cmn}{cm$^{-3}$}
\newcommand{\msun}{M$_{\odot}$}

\newcommand{\lam}{$\lambda$}

\newcommand{\lya}{\mbox{Ly$\alpha$}}
\newcommand{\lyb}{\mbox{Ly$\beta$}}
\newcommand{\lyg}{\mbox{Ly$\gamma$}}
\newcommand{\lyd}{\mbox{Ly$\delta$}}
\newcommand{\hi}{\mbox{H\,{\sc i}}}
\newcommand{\hii}{\mbox{H\,{\sc ii}}}

\newcommand{\heii}{\mbox{He\,{\sc ii}}}

\newcommand{\civ}{\mbox{C\,{\sc iv}}}
\newcommand{\ciii}{\mbox{C\,{\sc iii}}}
\newcommand{\cii}{\mbox{C\,{\sc ii}}}

\newcommand{\siiv}{\mbox{Si\,{\sc iv}}}

\newcommand{\siii}{\mbox{Si\,{\sc ii}}}
\newcommand{\siv}{\mbox{S\,{\sc iv}}}

\newcommand{\nv}{\mbox{N\,{\sc v}}}

\newcommand{\ovi}{\mbox{O\,{\sc vi}}}

\newcommand{\oi}{\mbox{O\,{\sc i}}}

\newcommand{\feii}{\mbox{Fe\,{\sc ii}}}
\newcommand{\feiii}{\mbox{Fe\,{\sc iii}}}

\newcommand{\mgii}{\mbox{Mg\,{\sc ii}}}

\newcommand{\aliii}{\mbox{Al\,{\sc iii}}}
\newcommand{\alii}{\mbox{Al\,{\sc ii}}}
\newcommand{\pv}{\mbox{P\,{\sc v}}}

\title[Structure and energetics of quasar outflows]{On the structure and energetics of quasar broad absorption-line outflows}
\author[F. Hamann, et al.]{Fred Hamann$^1$, Hanna Herbst$^{2}$, Isabelle Paris$^3$, Daniel Capellupo$^{4,5}$\\
$^{1}$Department of Physics \& Astronomy, University of California, Riverside, CA 92521, USA\\
$^{2}$Department of Astronomy, University of Florida, Gainesville, FL 32611, USA\\
$^3$Aix Marseille Universite, CNRS, LAM, Laboratoire d'Astrophysique de Marseille, F-13388 Marseille, France\\
$^4$Department of Physics, McGill University, Montreal, QC H3A 2T8, Canada\\
$^5$McGill Space Institute, McGill University, Montreal, QC H3A 2A7, Canada}
\begin{document}

\date{}

\pagerange{\pageref{firstpage}--\pageref{lastpage}} \pubyear{2018}

\maketitle

\label{firstpage}

\begin{abstract}
Quasar accretion-disk outflows might play an important role in galaxy evolution, but they are notoriously difficult to study due to line saturation and blending problems in the \lya\ forest. We circumvent these problems by constructing median composite spectra of diverse broad absorption lines (BALs) and `mini-BALs' in SDSS-III BOSS quasars at redshifts $2.3\leq z\leq3.5$. Sorting by \civ~\lam1549,1551 absorption-line strength with \aliii~\lam1855,1863 as an additional indicator of low ionisations (LoBALs) we find the following: (1) Deeper and broader BALs are accompanied by weaker \heii~\lam1640 emission lines, consistent with softer ionising spectra producing more effective radiative acceleration. (2) \pv~\lam1118,1128 absorption is present with resolved $\sim$1:1 depth ratios in all composites from mini-BALs to strong BALs indicating that line saturation, large total column densities $\log N_H(\textrm{cm}^{-2})\gtrsim22.7$, and large ionisation parameters $\log U\gtrsim-0.5$ are common. (3) Different observed depths in saturated lines identify inhomogeneous partial covering on spatial scales $\lesssim$0.006~pc, where weak/low-abundance transitions like \pv\ form in small high-column density clumps while stronger/broader lines like \civ\ form in larger regions. (4) Excited-state \siv*~\lam1073 and \ciii*~\lam1176 lines in BAL outflows indicate typical densities $n_e\gtrsim3\times10^5$~\cmn\ and maximum radial distances $R\lesssim23$~pc from the quasars. (5) For reasonable actual distances, the median BAL outflow has \textit{minimum} kinetic energy $L_K/L\gtrsim0.005(R/1.2\textrm{pc})$ sufficient for feedback to galaxy evolution. (6) LoBAL quasars have the largest median outflow column densities, highest velocities, and weakest \heii~\lam1640 emission in our study; they appear to be at one extreme in a distribution of quasar properties where softer ionising spectra drive more powerful outflows. \\ 

\vspace{-0.05cm}
\vspace{0.07cm}
\end{abstract}


\section{Introduction}
Accretion-disk outflows from quasars might provide important kinetic-energy feedback to their host galaxies that disrupts star formation and suppresses  accretion onto the central black hole \citep[e.g.,][and refs. therein]{DiMatteo05, Hopkins08, Fabian12, Liu13, Weinberger18, Carniani16, Carniani17, Rupke17}. The effectiveness of quasar feedback depends on the kinetic energy yields and the efficiency with which the outflows couple to gas in the host galaxies. These are complex issues not fully understood, but one theoretical study indicates that important feedback effects require outflow kinetic energy luminosities relative to the quasar bolometric of $L_K/L\sim 0.005$ \citep{Hopkins10}. 

Quasar outflows are often studied via blueshifted broad absorption lines (BALs) in rest-frame UV spectra. BALs are generally defined to have full widths at half minimum FWHM $\gtrsim$ 2000 km s$^{-1}$ at velocity shifts $>$3000 \kms\ in the quasar rest frame \citep{Weymann91}. `Mini-BALs' are more loosely defined to be narrower than BALs but still broad enough to clearly identify quasar-driven outflows \citep[][Chen et al., in prep.]{Hamann04, Paola08, Moravec17}. Their FWHMs extend down to a few hundred km s$^{-1}$. Quasar outflows are also classified by the presence or absence of low-ionisation absorption lines. BALs and mini-BALs generically have high-ionisation lines such as \civ\ \lam 1549,1551, while `LoBALs' are a subset of BALs that have, in addition to the high-ionisation BALs, strong low-ionisation lines such as \mgii\ \lam 2798,2803, \cii~\lam 1335, and \aliii~\lam 1855,1863 \citep[e.g.,][and refs. therein]{Trump06}. 

One of the key problems in quasar outflow studies is understanding the physical relationships between the different observed line types: mini-BALs, BALs, and LoBALs. Studies that show dramatic changes of one outflow line type into another \citep{Leighly09, Hall11, Paola13, FilizAk13, Rogerson16, Moravec17}, or presenting a detailed analysis of diverse lines in individual quasars \citep[e.g.,][Chen et al., in prep.]{Moe09, Misawa14, Misawa14b, Moravec17}, suggest that the different outflow line types are part of a single general outflow phenomenon viewed at different times or different orientations \citep[see also][]{Ganguly01, Elvis00, Hamann04, Elvis12, Hamann12, Matthews16}. LoBALs are a particularly important outflow type because they are more common in red quasars with, perhaps, higher accretion rates and, therefore, they might uniquely identify vigorous black hole accretion and outflows during early/active/obscured stages of quasar/host-galaxy evolution \citep{Voit93, Urrutia09, Lazarova12, Urrutia12, Glikman12}

A more basic problem is determining the outflow physical properties, such as the column densities and kinetic energies. Attempts to measure these properties from observed absorption lines are often hampered by saturation effects and line blending problems in the \lya\ forest (at quasar-frame wavelengths $\lesssim$1216 \AA). Blending in the forest is more problematic at high redshifts corresponding to the peak periods of quasar activity and massive galaxy formation. Column densities, in particular, are notoriously difficult to measure because line saturation can be masked in observed spectra by partial covering of the background light source along our lines of sight \citep{Barlow97, Hamann97, Hamann98, Arav99, Arav99b}. Saturated absorption lines can be weak (not reaching zero flux) because unabsorbed light fills in the bottoms of the line troughs. A further complication is that the outflows appear to be spatially inhomogeneous, so they present a distribution of column densities (and line optical depths) across the projected area of the quasar emission source \citep[][]{deKool02, Hamann01, Hamann04, Arav05, Arav08, Arav13}. This leads to optical depth-dependent covering fractions and thus optical depth-dependent line strengths in observed spectra, even in cases where all of the lines are optically thick \citep[see also][Hamann et al., submitted, Chen et al., submitted]{Moravec17}. 

Line saturation and partial covering can be diagnosed easily if the lines are narrow enough to resolve the individual components in doublets such as \siv~\lam 1394,1403 and \civ~\lam 1548,1551 \citep{Hamann97,Barlow97}. These doublets and others in the rest-frame near-UV have intrinsic optical depth ratios of $\sim$2:1 (for the short:long wavelength components). Observed line depth ratios near 1:1 indicate optical depths $\tau \gtrsim 3$ in the weaker doublet component. Observed ratios between 1:1 and 2:1 are more ambiguous; they can be caused by moderate line optical depths (of order unity) or by saturated absorption ($\tau \gg 1$) in an inhomogeneous medium with optical depth-dependent covering fractions. Thus the interpretation of observed line ratios, and the physical meaning of any single value of the ion column density deduced for an inhomogeneous absorber, require some knowledge or assumptions about the small-scale spatial structure of the outflows \citep[e.g., the column density spatial distribution,][]{deKool02, Hamann04, Arav08}. Another challenge is line blending; the doublet analysis described above is often not feasible for BALs because the lines are too broad and blended to resolve the separate components. For example, the \civ\ doublet separation, 498 \kms, is less than BAL line widths $\gtrsim$2000 \kms . 

Another way to diagnose line saturation and constrain the column densities in complex partial covering situations is to search for low-abundance absorption lines like \pv\ \lam 1118,1128. \pv\ has ionisation requirements very similar to \civ , so the ions should coexist spatially in quasar outflows. However, their abundance ratio in the Sun is C/P $\sim$ 1000 \citep{Asplund09} and photoionisation models based on solar abundances predict that the ratio of \civ /\pv\  line optical depths should be in the range of several hundred to $\sim$1200 \citep[][and Hamann et al., in prep.]{Hamann98, Leighly09, Leighly11, Borguet12, Chamberlain15}. The presence of \pv\ absorption lines therefore implies that \civ\ and many other common outflow lines such as \siiv, \nv~\lam 1239,1243, and \ovi~\lam 1032,1038 are highly saturated. It also requires large \textit{minimum} total column densities that are conservatively $\log N_{H}(\textrm{cm}^{-2}) \gtrsim 22$ (depending on line strengths and other physical parameters, again for solar abundances) \textit{and} large ionisation parameters to photoionise these large columns of \pv-absorbing gas (see refs. above). 

The detection rate of \pv\ absorption lines in quasar outflows has not been established due primarily to blending problems in the \lya\ forest. \cite{Capellupo17} identified a sample of 167 BAL quasars with definite or probable \pv\ absorption based on visual inspection of spectra in the Baryon Oscillation Spectroscopic Survey \citep[BOSS,][]{Dawson13,Paris17} of the Sloan Digital Sky Survey III \citep[SDSS-III,][]{Eisenstein11}. Some of the quasars in that study have clearly-resolved \pv\ doublets troughs that reveal $\sim$1:1 depth ratios, such that the \pv\ itself is saturated and the total outflow column densities could be much larger than the lower limit mentioned above. \cite{Moravec17} presented a detailed analysis of a single well-measured BAL quasar where a $\sim$1:1 doublet ratio in \pv\ indicates total column densities $\log N_{H}(\textrm{cm}^{-2}) \gtrsim 22.4$ and a kinetic energy luminosity, $L_K/L\gtrsim 0.0065$, which is sufficient for feedback to galaxy evolution (if the outflow is located roughly $R\sim 2$ pc from the quasar based on their line variability estimates). Analysis by \cite{Chamberlain15} of another well-measured \pv\ BAL quasar indicates similar total column densities but a much larger kinetic energy based on their estimates of a much larger radial distance of this outflow from the quasar.

In the present study, we construct median composite spectra of BAL and mini-BAL quasars at redshifts $2.3<z<3.5$ selected from SDSS-III BOSS. We sort the quasars by a variety of outflow line properties, and we include the sample of \pv-selected BAL quasars from \cite{Capellupo17} for additional comparisons. We show that composite spectra are extremely valuable to estimate the typical physical properties of quasars and to search for trends across the quasar samples. Composite spectra also provide an enormous advantage over individual quasar spectra by improving the signal-to-noise ratios and `averaging out' of the Ly$\alpha$ forest absorption lines to reveal the full range of diagnostic outflow lines available in the near-UV. For each quasar subsample, we construct composite spectra in the quasar rest frame to examine the emission-line properties and outflow velocity distributions, and in the absorber frame to study the the absorption-line strengths and profiles. This work builds upon the recent studies of BAL quasar composite spectra by \cite{Baskin13} and \cite{Baskin15}.  

Section 2 below presents an overview of the BOSS quasar sample and the selection criteria we used to define the outflow subsamples. Section 3 describes the procedures for constructing the composite spectra. Section 4 presents the composite spectra and the main results. Section 5 provides a discussion with further analysis of the broader implications of our study for quasar outflows and feedback to galaxy evolution. Section 6 gives a brief summary. 

\section{Quasar Samples \& BOSS Spectra}

We select quasars from the SDSS-III BOSS quasar catalog for data release 12 \citep[hereafter DR12Q,][]{Paris17}. The BOSS spectra cover observed wavelengths from $\sim$3600 \AA\ through $\sim$10,000 \AA\ at resolutions of $\lambda/\Delta\lambda \sim 1300$ at the blue end to $\sim$2600 in the red \citep{Dawson13}. We require that the spectra have signal-to-noise ratios $>$ 2 at rest-frame 1700 \AA\ (i.e., \texttt{snr\_1700} $>$ 2 in DR12Q) and that the quasars have redshifts in the range $2.3 \leq z \leq 3.5$. The redshift range ensures that the BOSS spectra cover rest wavelengths from at least $\sim$1090 \AA\ to $\sim$2250 \AA\ in every quasar, which encompasses important lines from \pv\ \lam 1118,1128 to \civ\ \lam 1548,1551 and \aliii\ \lam 1855,1863. The maximum redshift is also useful to exclude high-$z$ quasars with severe \lya\ forest contamination (at wavelengths $\lesssim$1216\AA\ in the quasar frame). The result is a full sample of 104,433 quasars. Roughly two thirds of them have coverage across \ovi\ \lam 1032,1038 and half cover \mgii\ \lam 2798,2803. 

BOSS spectra have known flux calibration problems related to differential atmospheric refraction and guiding offsets. We apply flux corrections to every spectrum according to the prescription developed by \cite{Harris16}, which depends on the airmass during the observations. These corrections are approximate for individual quasars, but they work well on average for larger samples and so they are appropriate for our analysis of composite quasar spectra. We also note that none of our results depend on the accuracy of the flux corrections. 


From the full quasar sample described above, we create subsamples sorted by outflow line properties. The line parameters we use to sort the BAL and mini-BAL subsamples are the \civ\  balnicity index (BI) and absorption index (AI) as listed in DR12Q \citep{Paris17}. BI is a modified equivalent width that measures absorption at blueshifted velocities between 3000 to 25,000 \kms\ if it has a contiguous velocity width $>$2000 \kms\ at depths $>$10 percent below the underlying emission spectrum \citep{Weymann91}. AI is similar but it includes narrower features $>$450 km s$^{-1}$ wide at blueshifted velocities from 0 to 25,000 \kms\ \citep{Hall02}. We require that the recorded BI and AI values have signal-to-noise ratios $>$4 (specifically \texttt{bi\_civ}/\texttt{err\_bi\_civ} $> 4$ and \texttt{ai\_civ}/\texttt{err\_ai\_civ} $> 4$ as listed in DR12Q). We also require that BAL quasars with BI $>$ 0 are also identified as BALs by visual inspection (e.g., \texttt{bal\_flag\_vi} = 1). The total number of BAL quasars selected this way is 6856. 

Figure 1 plots the redshift and the apparent and absolute $i$-band magnitude distributions of the full quasar sample, the BAL quasars defined above, and a subset of non-outflow quasars defined by the absence of any broad absorption features defined by BI = 0, AI = 0, and \texttt{bal\_flag\_vi} = 0. The redshift distributions of these subsamples are very similar, each with median values near $\left<z\right>\sim 2.6$. The magnitude distributions are offset in the sense that BAL quasars in our study tend to be brighter and more luminous than non-outflow quasars. The median magnitudes are $\left<M_i\right> = -25.72$, $-$25.59, and $-$26.30, and $\left<i\right> = 20.49$, 20.60, and 20.06 for the full sample, non-outflow quasars, and BAL quasars, respectively. 

\begin{figure}
\includegraphics[width=0.99\columnwidth]{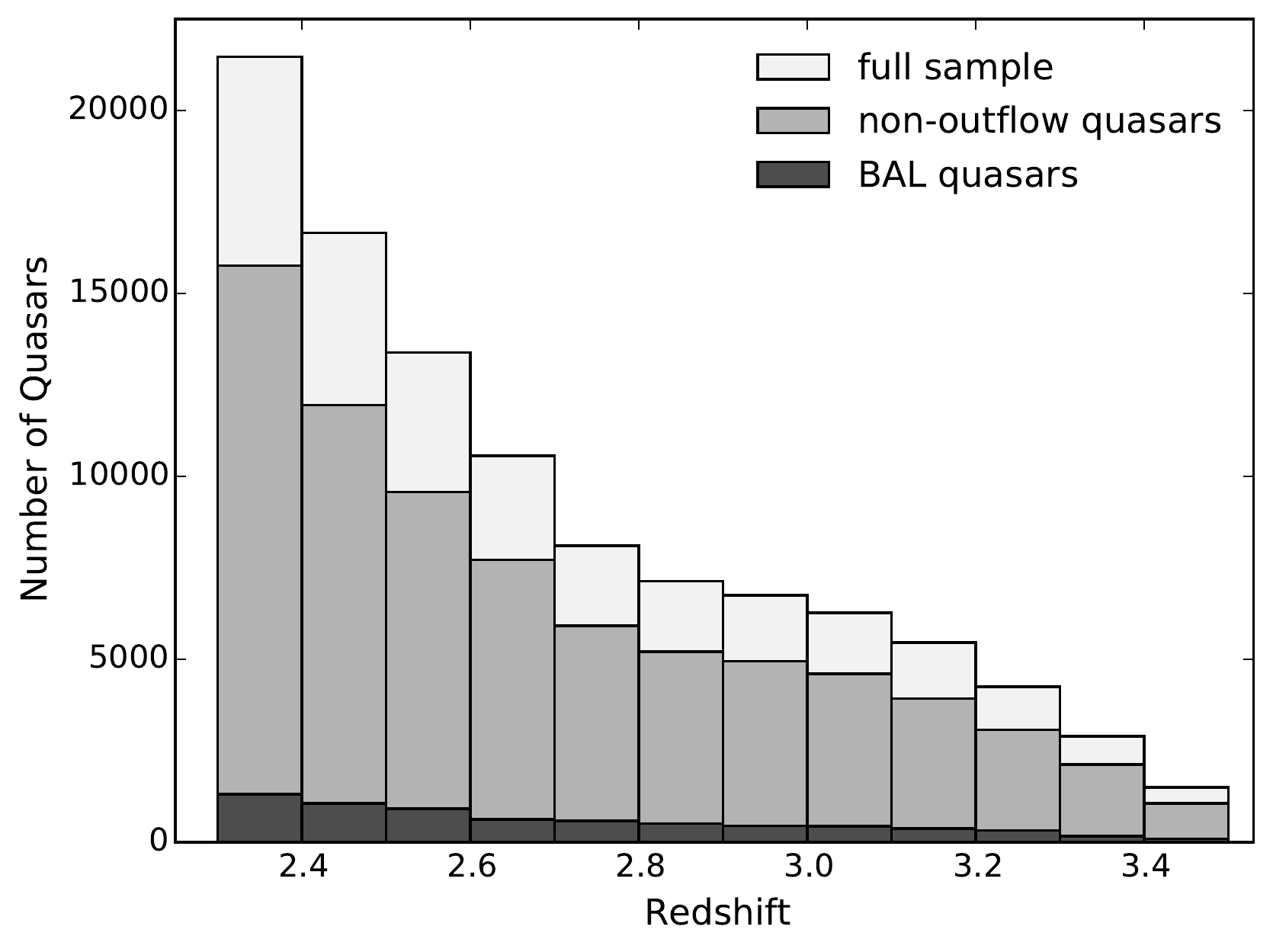}
\includegraphics[width=0.99\columnwidth]{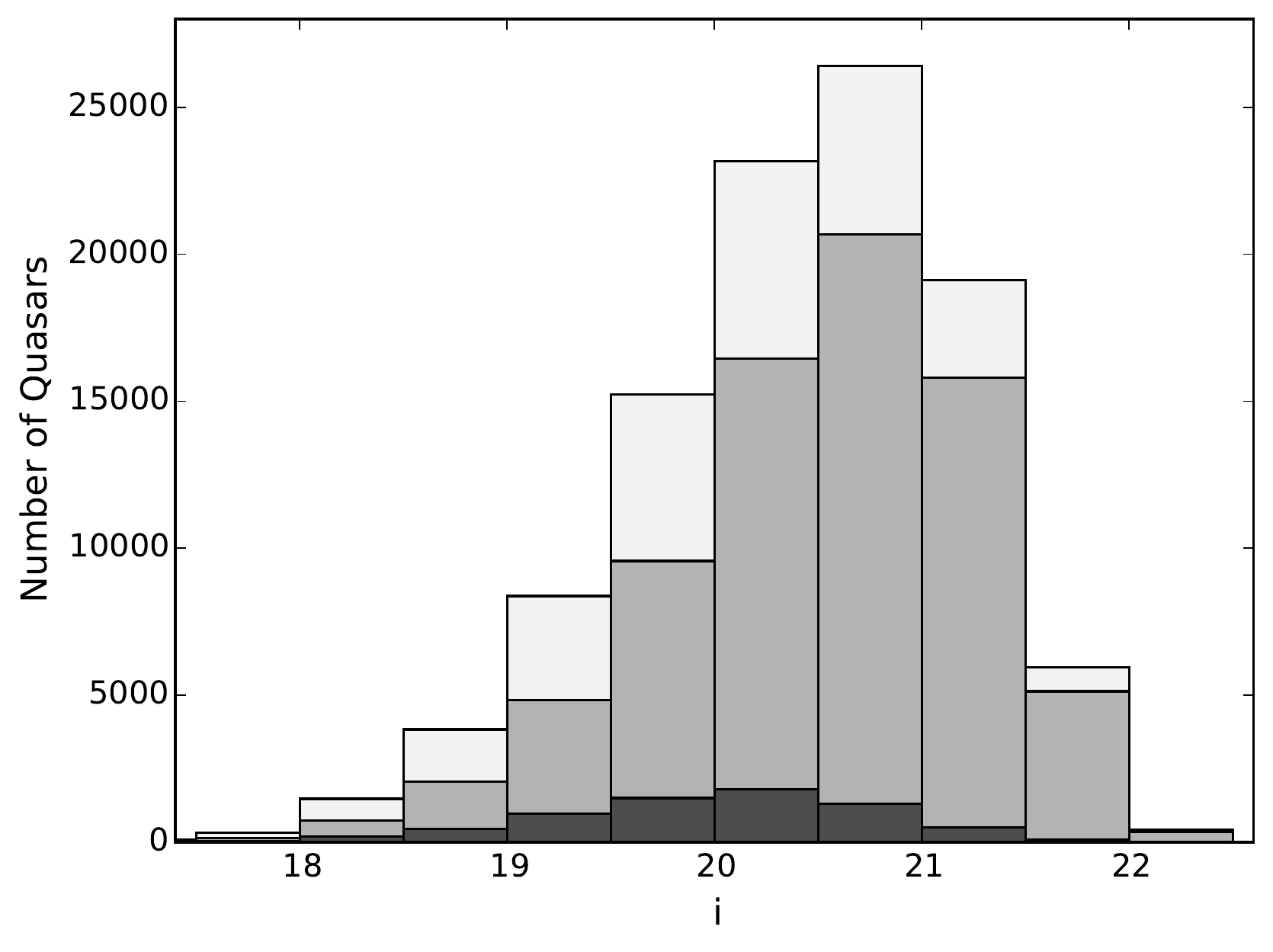}
\includegraphics[width=0.99\columnwidth]{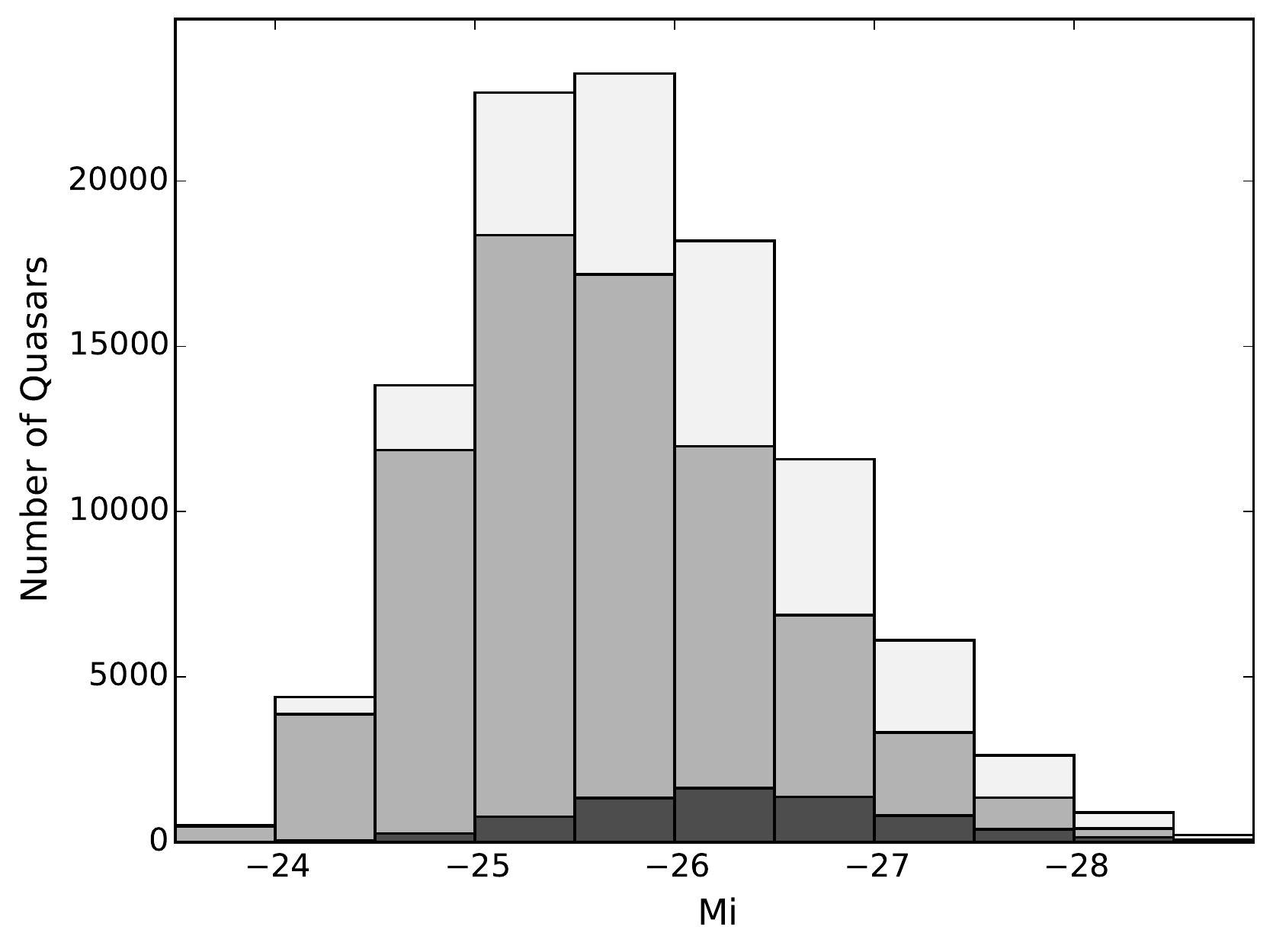}
\vspace{-8pt}
 \caption{Distributions in redshift and in apparent and absolute ($M_i$) $i$-band magnitudes for the full quasar sample, non-outflow quasars, and BAL quasars in our study. The redshift distributions of these samples are very similar, but their $i$ and $M_i$ distributions differ in the sense that the BAL quasars tend to be brighter and more luminous (see Section 2.0).}
\end{figure}

The $i$-band magnitudes measure rest wavelengths near $\lambda \sim 2100$ \AA\ at the median redshift of our study. Thus the median apparent magnitudes correspond to median specific luminosities of $\lambda L_{\lambda}(2100\textrm{\AA}) \approx 5.3\times 10^{45}$ erg~s$^{-1}$ and $\lambda L_{\lambda}(2100\textrm{\AA}) \approx 7.8\times 10^{45}$ erg~s$^{-1}$ for the full sample and BAL quasars, respectively (using the online cosmology calculator developed by \cite{Wright06} with parameters $H_o = 69.6$ \kms\ Mpc$^{-1}$, $\Omega_M = 0.286$, and $\Omega_{\Lambda}=0.714$ from \cite{Bennett14}). Adopting the bolometric correction $L=3.9\lambda L_{\lambda}(2100\AA)$ based on a standard quasar spectral energy distribution \citep[SED, e.g.,][and refs. therein]{Hamann13} then yields median bolometric luminosities of $L\approx 2.1\times 10^{46}$ erg~s$^{-1}$ for the full sample and $L\approx 3.0\times 10^{46}$ erg~s$^{-1}$ for the BAL quasars.

Table 1 lists all of the outflow subsamples that we consider in our study. They are grouped in the table by common properties and sorting parameters. Table 1 lists specifically the group names (such as `all BALs'), the parameter values (such as `BI') used to define/sort the subsamples, and then the numbers of quasars (\# QSOs), the median absolute $i$-band magnitudes (Med $M_i$), and the median balnicity index (Med BI) in each subsample. Every outflow subsample (with data listed under the generic column heading `BAL') is accompanied by a much larger non-outflow subsample (under the heading `non-BAL') that we use to remove the broad emission lines from the outflow composite spectra. The non-outflow quasars are required to have BI = 0, AI = 0, and \texttt{bal\_flag\_vi} = 0 plus additional constraints for the emission-line matching to the outflow subsamples (as described in Section 3). The subsections below provide more information on the quasar subsamples. 

\begin{table}
 \centering
  \caption{Summary of the outflow and matched non-outflow subsamples. The column headings are: Sample = outflow group name and sorting parameters, \# QSOs = number of quasars in each bin; Med M$_{i}$ = median absolute $i$-band magnitude; Med BI = median \civ\ balnicity index. Subheadings BAL and non-BAL refer generically to the outflow and non-outflow subsamples, respectively.}
  \tabcolsep=0.14cm
  \begin{tabular}{r c c c c c}
  \hline
   & \multicolumn{2}{c}{--- \# QSOs ---} & \multicolumn{2}{c}{--- Med $M_{i}$ ---} & Med BI \\
  Sample & BAL & non-BAL & BAL & non-BAL & BAL\\
\hline
\underbar{mini-BALs:} & 1056 & 12672 & -25.93 & -25.90 & 0 \\
 \hline
 \underbar{all BALs:} & \\
 0$<$BI$<$500 & 1984 & 23808 & -26.26 & -26.13 & 194 \\
 500$<$BI$<$1000 & 1005 & 15075 & -26.20 & -26.13 & 730 \\
 1000$<$BI$<$2000 & 1297 & 18158 & -26.17 & -26.14 & 1449 \\
 2000$<$BI$<$5000 & 1892 & 22704 & -26.33 & -26.18 & 3079 \\
 BI$>$5000 & 678 & 10170 & -26.68 & -26.29 & 6499\vspace{-6pt}\\
 \multicolumn{6}{r}{.................................................................................................................}\\
500$<$BI$<$5000 & 4194 & 41940 & -26.25 & -26.08 & 1808 \\
\hline
\underbar{LoBALs:} & \\
REW(\aliii )=0$^a$ & 3147 & 25176 & -26.55 & -26.34 & 2118 \\
2$<$REW(\aliii )$<$6 & 146 & 11680 & -26.77 & -26.42 & 4401 \\
6$<$REW(\aliii )$<$10 & 92 & 7360 & -26.69 & -26.41 & 4637 \\
REW(\aliii )$>$10 & 147 & 5145 & -26.96 & -26.30 & 6686\vspace{-6pt}\\
 \multicolumn{6}{r}{.................................................................................................................}\\
6$<$REW(\aliii )$<$24 & 214 & 6420 & -26.82 & -26.32 & 5582 \\
\hline
\underbar{HiBALs:} & \\
BI$>$5000 & 474 & 7110 & -26.57 & -26.17 & 6209 \\
\hline
\underbar{\pv -selected BALs:} & \\
all BALs: matched BI& 1490 & 14900 & -26.45 & -26.06 & 3928 \\
definite \pv& 81 & 6480 & -26.41 & -26.15 & 4038 \\
all \pv& 167 & 8350 & -26.43 & -26.15 & 3801 \\
\hline
\hline
\end{tabular}
\begin{tablenotes}
\item $^a$This is a HiBAL sample selected to have \texttt{snr\_1700} $>$ 5 and BI $>$ 500 like the LoBALs in this group (Section 2.3).
\end{tablenotes}
\end{table}

\subsection{All BALs}  

We divide the 6856 BAL quasars in our study into five subsamples sorted by BI, ranging from weak \civ\ BALs with 0 $<$ BI $<$ 500 to the strongest BALs with BI $>$ 5000. We refer to this BAL quasar group as `all BALs' because, unlike the LoBAL and HiBAL subsamples described below, it includes all BALs with the requisite BI values. We also consider a larger subsample of BALs with 500 $<$ BI $<$ 5000 to create a single best high signal-to-noise composite BAL spectrum based on 4194 quasars. The BI range of this subsample is designed to exclude weak or questionable BALs at small BI and very broad BALs at large BI that cause line blending and blurring in the composites (Section 3). 

\subsection{Mini-BALs}

We construct a mini-BAL subsample that requires BI = 0, AI $>$ 1000, and a minimum outflow velocity $>$1000 km s$^{-1}$ (as specified by \texttt{vmin\_450} $>$ 1000 in DR12Q). Our goal is to capture outflow lines that are narrower and weaker than traditional BALs while still avoiding narrow absorption lines that might not form in quasar outflows. BI = 0 ensures that the mini-BALs in this subsample do not satisfy the standard definition of a BAL with \civ\  FWHMs $\gtrsim 2000$ at velocity shift $>$ 3000 \kms\ \citep{Weymann91}. The minimum velocity shift avoids a large number of low-speed `associated' absorption lines (AALs) that can be broadened by blends to satisfy the minimum AI threshold. These systems have deep square absorption troughs that do not resemble BALs and do not clearly form in quasar-driven outflows. An example of the type of low-speed AAL system we aim to avoid is described by \cite{Hamann01}. The minimum AI threshold is essential to exclude narrow \civ\  absorption lines at all velocity shifts that are unrelated to the quasars. These systems are the main source of contamination for mini-BAL subsamples defined by AI. 

We experimented with different minimum thresholds in AI and velocity shift and checked the results by visually inspecting the individual BOSS spectra. We find that the main source of contamination is narrow lines that belong to damped Ly$\alpha$ (DLA) or strong Lyman limit systems. We reject all quasars with DLAs based on our visual inspections \citep[aided by a preliminary DLA catalog kindly provided to us by Pasquier Noterdaeme, see also][]{Noterdaeme12}. However, these rejections are insufficient because they  are limited to strong DLAs with \lya\ in the BOSS wavelength coverage. Thus we additionally reject quasars that have a combination of absorption-line properties like DLAs and Lyman-limit systems, namely: strong Ly$\alpha$ absorption that reaches zero intensity, narrow \civ\  line profiles with well-resolved doublets and square profiles that do not appear BAL-like, and strong narrow low-ionisation lines such as \oi\ \lam 1304, \cii\ \lam1335, \feii\ \lam 1608, \siii\ \lam 1260, and \siii\ \lam 1527. We keep quasars if do not clearly match these rejection criteria. Our visual inspections of these ambiguous cases that might have unrelated absorption lines instead of outflow mini-BALs account for $\lesssim$10 percent of our final sample of 1056 mini-BAL quasars.  

\subsection{LoBALs and HiBALs} 

We create subsamples of low-ionisation BAL quasars (LoBALs) sorted by the \aliii\ 1855,1863 \AA\ absorption-line rest equivalent width (REW) listed in DR12Q. We use \aliii\ instead of the more standard definition of LoBALs based on MgII 2796, 2803 \AA\ absorption \citep[e.g.,][]{Trump06} because \aliii\ is within the BOSS spectral coverage for every quasar and its REW is conveniently listed in DR12Q. and it is known to be a good tracer of low-ionisation gas \citep{Reichard03}. The resulting composites and our inspections of the individual quasar spectra both confirm that \aliii\ is a reasonable surrogate for \mgii\ because stronger \aliii\ absorption is accompanied by stronger absorption in \mgii\ and other lower ions \citep[Section 4, also][]{Reichard03}. 

We divide the LoBALs into three intervals of REW(AlIII) plus a larger subsample with $6<\textrm{REW(\aliii)}<24$ \AA\ that includes most of the moderate-to-strong LoBALs (see Table 1). Note that \aliii\ REWs are available in DR12Q only for BAL quasars with BI $>$ 500 and median signal-to-noise ratios $>$ 5 near 1700 \AA\ rest (e.g., \texttt{snr\_1700} $>$ 5 in DR12Q. For comparison to these LoBAL quasars, we create another subsample of high-ionisation BALs (HiBALs) that satisfy the same requirements for BI $>$ 500 and \texttt{snr\_1700} $>$ 5, but \textit{without} significant \aliii\ based on REW(\aliii) = 0 recorded in DR12Q. 

We created a full series of HiBAL subsamples sorted by BI analogous to the `all BALs' in Section 2.1 but with LoBAL quasars removed based on REW(\aliii ) $>$ 0. However, only the largest BI bin with BI $>$ 5000 has enough LoBALs discarded to be significantly different than its corresponding `all BAL' subsample. Therefore, we keep only this one HiBAL subsample with BI $>$ 5000 for consideration (Table 1).    

\subsection{PV-selected BALs}

Finally, we include the samples of \pv\ BAL quasars discovered by \cite{Capellupo17} visual inspection the of 2694 BAL quasars at redshifts $2.3 < z < 4.5$ in BOSS data release 9 \citep{Paris12}. They found specifically 81 quasars with `definite' \pv\ BALs and 86 quasars with `probable' \pv\ BALs. They note that their `probable' classification is conservative such that most or all of the quasars with that designation have real \pv\ absorption lines. We create composite spectra for definite \pv\ BALs and their combined definite+probable sample, which we list as `def \pv ' and `all \pv ', respectively, in Table 1. These samples underrepresent the true numbers of \pv\ BAL quasars in BOSS by potentially large factors due to the difficulties of securely identifying \pv\ absorption in the \lya\ forest. Strong \pv\  absorption is generally required for secure detections in individual, which leads to a bias toward strong \pv\ absorption and probably strong BALs overall in these PV-selected samples. 

\cite{Capellupo17} already presented composite spectra of their PV-selected BAL quasars in the absorber frame defined by an average of the minimum and maximum velocities in the BI integration (\texttt{vmin\_civ\_2000} and \texttt{vmax\_civ\_2000} in the BOSS quasar catalog). We construct new composite spectra of these same PV-selected quasars to update their results in two important ways. First, we define the absorber-frame redshifts more precisely at the absorption minimum in the \civ\ troughs, which reduces velocity uncertainties (blurring) in the composites to provide more accurate measures of the median BAL profiles. Second, we examine additional comparison samples including one with BAL quasars matched in BI to the PV-selected BALs (listed as the `matched BALs' in Table 1). This comparison at a given \civ\  BAL strength (BI) is valuable to study the absorption-line features and outflow physical properties that correlate specifically with strong \pv\ BALs. 

\section{Composite spectra}

We create median composite spectra in both the quasar frame and in the absorber frame for all of the subsamples listed in Table 1. The first step is to normalize the individual quasar spectra by dividing by the median flux in a small window free of spectral lines from 1690 to 1730 \AA . Then we combine the spectra using medians instead of means because the medians reject outliers and preserve the relative strengths of emission and absorption lines \citep{Vandenberk01}. We also find that the median composite spectra yield higher signal-to-noise ratios than composites computed using simple means or means computed with `sigma-clipping.' We experimented with different sigma-clipping thresholds to reject the top and bottom 20 percent, 30 percent, and 40 percent of flux values at each pixel in normalized spectra. Larger clipping thresholds produce both lower signal-to-noise ratios (due to smaller numbers of spectra contributing) and poorer results in the Ly$\alpha$ forest where large numbers of spectra are needed to average-out the forest lines. However, the important emission- and absorption-line behaviors discussed in Sections 4 and 5 are roughly the same, within uncertainties, in all of the median, mean, and sigma-clipped mean composite spectra. 

We construct quasar-frame composites by shifting the individual spectra to rest wavelengths defined by the principle component analysis (PCA) redshifts listed in DR12Q, which are estimated to have typical uncertainties of $\sim$500 km s$^{-1}$ (\citealt{Paris12}, but see also \cite{Shen16} and \cite{Denney16}). In rare cases where a PCA redshift is not available or the velocity difference between the PCA redshift and the visual-inspection redshift in DR12Q exceeds 5000 km s$^{-1}$, we use the visual inspection redshift instead. 

We construct absorber-frame composites by shifting to a rest frame defined by the wavelength of minimum flux in the \civ\  trough between the integration limits for BI (for BAL quasars) or AI (for mini-BALs), as provided in DR12Q (eg., \texttt{vmin\_civ\_2000} and \texttt{vmax\_civ\_2000} or \texttt{vmin\_civ\_450} and \texttt{vmax\_civ\_450}, respectively). We identify the minimum flux after smoothing the spectra by a Gaussian filter with standard deviations of roughly 400 km s$^{-1}$. Smoothing is essential to avoid noise spikes and deep narrow unrelated absorption lines in our determinations of the flux minima. The amount of smoothing is a compromise between being broad enough to avoid these contaminating features but narrow enough to not introduce redshift uncertainties that blur the final absorber-frame composite spectra. 

Each absorber-frame composite is divided by a composite spectrum of non-outflow quasars roughly matched to the outflow quasars in emission-line strength. The non-outflow quasars are required to have BI~=~0, AI~=~0, and \texttt{bal\_flag\_vi}~=~0 (as in Section 2.1). Dividing by the non-outflow composite removes broad emission-line features from the outflow composite to reveal better the outflow absorption lines. We follow the procedure in \cite{Baskin15} to match quasars using the HeII 1640 \AA\ emission-line strength measured approximately from the median flux near $\sim$1640 \AA\ relative to the continuum at $\sim$1700 \AA . We additionally require that the matched quasars lie within a range of absolute $i$-band magnitudes ($M_i$ in DR12Q) similar to the outflow quasars. This helps to minimize differences in the emission-line strengths due to the Baldwin Effect, which describes an empirical relation between the quasar luminosities and the emission-line REWs \citep{Baldwin77, Kinney90, Dietrich02, Shields07}. The matching based on $M_i$ is particularly helpful for LoBAL and other strong-BAL quasars (large BI) that have very weak or undetected \heii\ emission lines (see Figures 2 and 3 below). Numerous trial-and-error experiments show that quasars with strong BALs tend to have narrower emission-line profiles than non-outflow quasars at the same $M_i$. To compensate for this, we favor slightly fainter $M_i$ magnitudes in the matched non-outflow subsamples. Table 1 lists the numbers of quasars and the median $M_i$ for each pair of matched outflow/non-outflow subsamples. 

After matching, the last steps for the absorber-frame composites are to 1) randomly apply velocity shifts to the non-outflow quasar spectra using the same shift distribution as the outflow quasars, 2) compute a median spectrum from the shifted non-outflow quasar spectra, and 3) divide that result into the absorber-frame outflow composite. We will refer to these emission-line-corrected absorber-frame composite spectra simply as the absorber-frame composites.  

\section{Empirical Results}

\subsection{Quasar-frame composite spectra}

Figure 2 shows normalized quasar-frame composite spectra for most of the subsamples described in Section 2 and listed in Table 1. Figure 3 shows these same composites on an expanded scale at shorter wavelengths. These figures include a non-outflow composite spectrum (grey curves) constructed from 28,888 quasars having the same median absolute magnitude, $M_i = -26.23$, as the BAL quasars with $500<$~BI~$<5000$ in our study. The spectra in Figures 2 and 3 clearly illustrate the value of median composites for averaging over the Ly$\alpha$ forest to reveal intrinsic quasar spectral features. The only remaining effect of the forest is a roughly smooth suppression of the flux at wavelengths $<$1216 \AA , which should be the same for all of the quasar subsamples because they all have the same median redshift. The lower fluxes at $<$1216 \AA\ in the BAL quasars compared to the non-outflow composite are, therefore, a real effect caused by redder continuum slopes and (at shorter wavelengths and larger BI) overlapping BAL troughs. 

\begin{figure*}
\includegraphics[width=1.0\textwidth]{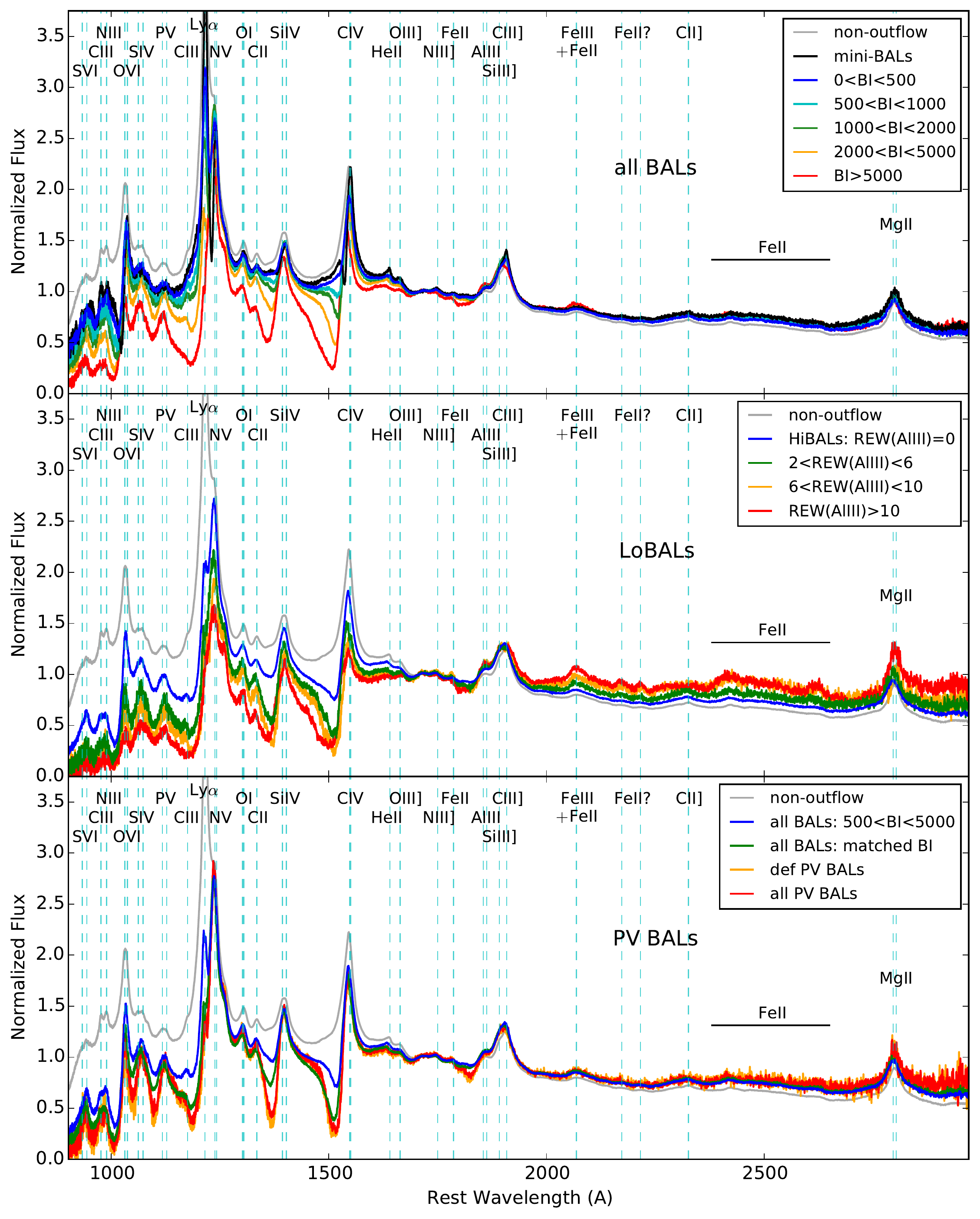}
\vspace{-16pt}
 \caption{Quasar-frame composite spectra normalized to unity at $\sim$1700 \AA. \textit{Top panel}: `all BAL' subsamples sorted by \civ\  balnicity index BI (Section 2.1). \textit{Middle panel}: three LoBAL composites compared to HiBALs defined by REW(\aliii ) = 0 (Section 2.3). \textit{Bottom panel}: PV-selected BAL quasars compared to two `all BAL' subsamples with $500<$~BI~$<5000$ or matched to the \pv -selected quasars in BI (Section 2.4). The grey curve in all three panels shows a non-outflow composite spectrum for quasars with median $M_i$ equal to the median for all BAL quasars. Rest wavelengths of broad emission lines are marked by dashed vertical lines with labels across the top in each panel. The horizontal bars near 2500 \AA\ mark a broad blend of \feii\ emission lines.}
\end{figure*}

\begin{figure*}
\includegraphics[width=1.0\textwidth]{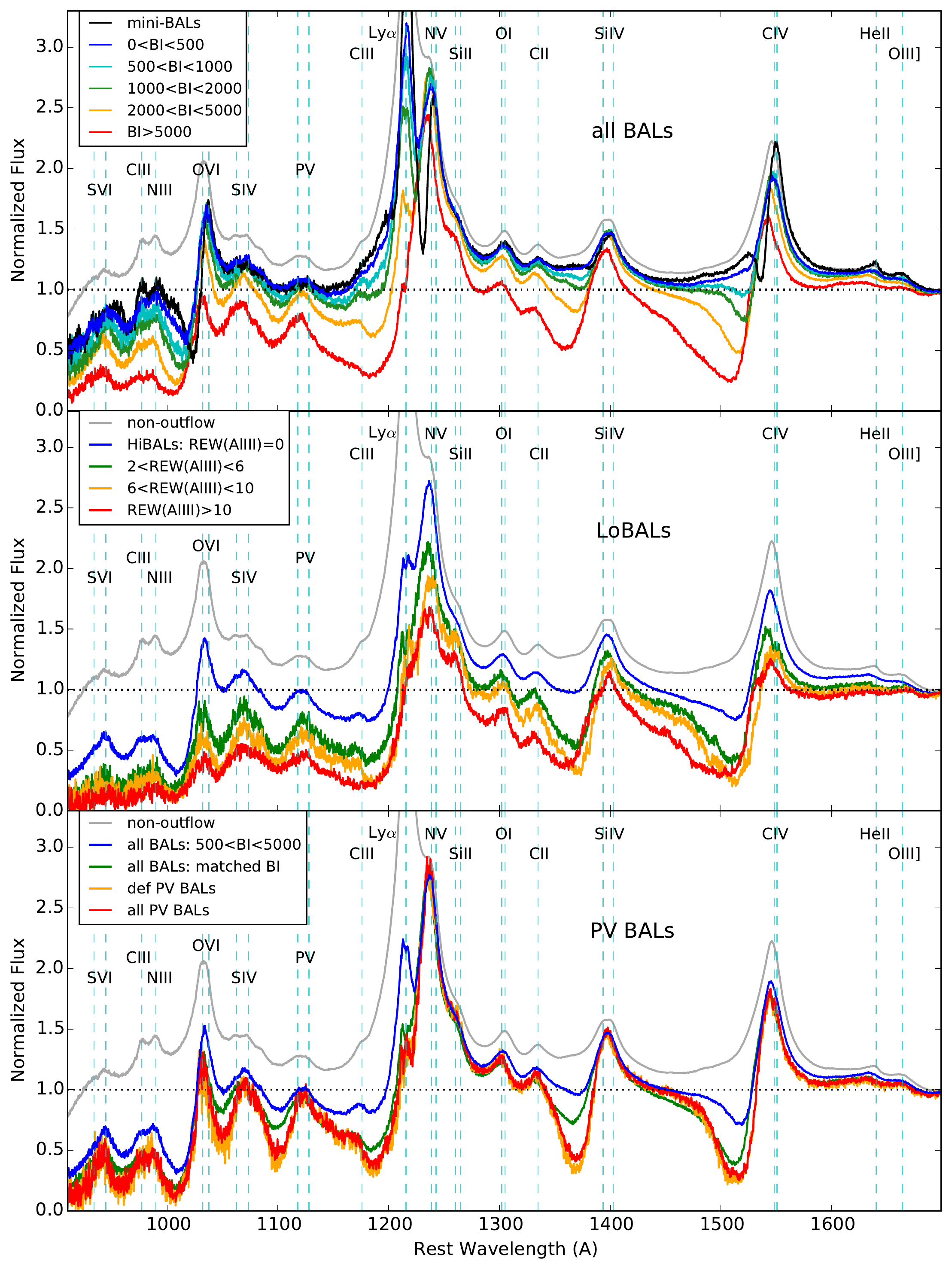}
\vspace{-18pt}
\caption{Expanded view of the quasar-frame composite spectra shown in Figure 2. Dotted horizontal lines at unit flux are drawn to guide the eye.}
\end{figure*}


\subsubsection{Near-UV continuum slopes \& luminosities}

All of the outflow subsamples plotted in Figures 2 and 3 have redder UV continua than typical non-outflow quasars at the same $M_i$ (shown by the grey curve). There is no obvious relationship of the continuum slope to \civ\ BAL strength (BI) in the `all BAL' samples (top panels). However, LoBAL quasars have markedly redder spectra than other BALs. These results are consistent with the quasar-frame composite spectra presented by \cite{Reichard03} and with many previous studies of quasar reddening \citep{Weymann91, Sprayberry92, Brotherton01, Richards03, Trump06, Knigge08, Dai08, Gibson09, Urrutia09, Scaringi09, Zhang10, Allen11, Pu14, Krawczyk15}. We will show in a forthcoming paper (Hamann et al., in prep.) that, among non-outflow quasars, redder UV spectral slopes also correlate with larger emission-line REWs. 


An important new result in our study is that BAL quasars tend to be brighter in $M_i$ by almost a factor of two compared to the non-outflow quasars (Figure 1). This might be a selection effect because the requirement for well-measured BI values in the BAL quasars tends to favor brighter quasars with higher signal-to-noise spectra in the BOSS archive. However, this selection bias should affect weak BALs more than strong ones, leading to preferentially brighter $M_i$ magnitudes at small BI. There is no trend like that in our samples. In fact, we find that the strongest BAL quasars with BI $>$ 5000 are the most luminous in $M_i$ (Table 1). Similarly, mini-BAL quasars with the weakest outflow lines tend to be less luminous than BAL quasars (by a factor of $\sim$2 compared to BAL quasars with BI~$>$5000). Thus the bias appears to be negligible. 

We conclude that there \textit{is} a significant trend for brighter $M_i$ magnitudes in quasars with stronger/broader outflow lines. A similar trend for larger maximum outflow velocities in more luminous quasars was discussed previously by \cite{Laor02}. Those authors argue that higher observed luminosities in BAL quasars, which they find to be accompanied by higher accretion rates (e.g., larger Eddington ratios, $L/L_{E}$), are expected results for BAL outflows driven by radiative forces. A more thorough discussion of this point for the quasars in our study will require future estimates of the black hole masses and $L/L_E$ ratios. 

Strong LoBAL quasars appear to be at an extreme in the observed luminosity trend. They have the brightest $M_i$ magnitudes of any of our subsample in spite of also having the reddest near-UV spectra. If the reddening is caused by dust obscuration, then the actual UV luminosities of strong LoBALs would be even larger compared to the other quasar subsamples. There is again some danger of a selection bias here because LoBAL classification depends on detections of \aliii\ absorption. However, the LoBAL accounting in DR12Q considers only high signal-to-noise spectra (Section 2.3). Within that select group, quasars with the strongest LoBALs (REW(\aliii) $>$ 10 \AA) are typically $\sim$1.5 times more luminous based on \textit{uncorrected} UV flux than non-LoBAL quasars with REW(\aliii) = 0 (Table 1). 

\subsubsection{Absorption-line trends}

Median absorption-line profiles in the quasar-frame composite spectra (Figures 2 and 3) represent the line profiles in individual quasars \textit{combined with} object-to-object differences in the line velocity shifts across the sample. The range of velocity shifts across the samples leads to median quasar-frame BAL profiles (Figure 3) considerably broader than typical individual BALs and, e.g., broader than the median BAL in the absorber frame (Figure 4 to 6 in Section 4.2 below).

Strong BALs with BI $>$ 5000 (top panel in Figure 3) have a median \civ\ profile with minimum flux at velocity v$\,\sim8000$ \kms\ and a blue wing that reaches at least $\sim$27,000 \kms\ (where it overlaps with the \siv\ \lam 1393,1403 emission line). The broadest quasar-frame \civ\ trough is in the strongest LoBAL composite defined by REW(\aliii)~>~10 \AA, with a flux minimum around v$\;\sim9000$ to 13,000 \kms\ and deeper absorption at high velocities than any other BAL composite. Weaker \civ\  BALs (smaller BI) have flux minima at progressively lower velocities with much less absorption at large velocities. For example, the moderate BALs with 1000~$<$~BI~$<$~2000 have median \civ\ flux minimum at $\sim$5000 \kms\ while the mini-BALs have minimum flux at only $\sim$2000 \kms . (Note, however, that the mini-BALs in our study have velocity shifts biased towards this value because we require v$\;> 1000$ \kms\ plus BI = 0, which implicitly avoids strong absorption at v$\; >3000$ \kms , Section 2.2.) 


The PV-selected BAL composites in the bottom panel of Figure 3 have a distinctly different appearance than the other composites. In particular, they have much deeper absorption not only in \pv\ but also in a variety of other lines including \siiv , \siv\ \lam 1063,1073, and \aliii\ compared to the `all BALs' sample matched in BI. The \pv-selected quasars also have \civ\  BAL profiles with a broad minimum near 8000 \kms, like other strong BALs matched in BI, but the troughs are deeper and slightly narrower with less absorption at large velocities. The narrower profiles in \pv-selected quasars result from a narrower object-to-object distribution in the velocity shifts. Nonetheless, it is an important result for the structure and energetics of quasar outflows that the high-column density gas traced by the low-abundance \pv\ line appears typically at velocities in the range $\sim$6000 to $\sim$10,000 \kms\ (Section 5.5). 

\subsubsection{Emission-line trends}

Figures 2 and 3 reveal several important emission-line behaviors. First, in the top panels, BAL quasars with weaker troughs (smaller BI) and smaller velocity shifts have progressively stronger emission lines in the higher ions, notably \civ\  and \heii\ \lam 1640, while their low-to-intermediate ionisation lines, such as the blend near \ciii] \lam 1909, remain roughly the same. The mini-BALs continue this trend with the weakest outflow lines (BI~=~0) accompanied by strong \civ\ and \heii\ emission lines similar to the non-outflow composite. LoBAL quasars are at the opposite extreme; they have the weakest \civ\ and \heii\ emission accompanied by \textit{enhanced} emission in low-ionisation lines like \feii\ and \mgii\ \lam 2796,2804. Inspection of Figure 3 shows, for example, that the normalized peak heights in \civ\ decrease from $\sim$2.2 in the non-outflow and mini-BAL composites to $\sim$0.6 in strong BALs with BI $>$ 5000 to $\sim$0.25 in the strongest LoBALs. The \heii\ emission line is more difficult to measure, but it appears to change roughly similarly by a factor of $\gtrsim$5 across this sequence -- being strongest in non-outflow and mini-BAL quasars to undetectable in the LoBAL composite. (Note that the weak emission `bump' just blueward of 1640 \AA\ in the strong BAL spectra is believed to be \feii\ emission, not \heii, e.g., \cite{Laor97}.) Emission-line trends like this have been noted before in BAL quasar studies \citep[e.g.,][]{Gibson09, Baskin13, Baskin15}. We discuss their physical implications in Section 5.1. 

\subsection{Absorber-frame composite spectra}

Figures 4 through 6 show absorber-frame composite spectra for all of the same outflow subsamples plotted in Figures 2 and 3. Figure 4 additionally shows the HiBAL subsample with BI~$>$5000 and LoBALs removed (Section 2.3). The rest wavelengths labeled in these plots are in the absorber frame. The emission lines are offset to longer wavelengths (by amounts indicated nominally by the offset between the emission and absorption lines in the quasar-frame Figures 2 and 3. Note that the apparent weak absorption dip in the BI~$>$~5000 composite at $\sim$1600 \AA\ is due to a slight mismatch between the emission line strengths in this BAL composite compared to its matched non-BAL composite (Section 3). 

Figure 7 plots some of the BAL profiles on a velocity scale measured from the flux minimum in \civ\ (Section 3). The profiles are normalized to unity in the continuum based on linear extrapolations between flux points on either side of the absorption troughs. This procedure can substantially underestimate the true continuum, and thus underestimate the true absorption strength, for \pv\ due to blending problems at those wavelengths. We do not attempt to correct for these underestimates. We note simply that they appear to be important for the strong-BAL composites with median BI $\gtrsim$ 3000.  

\begin{figure*}
\includegraphics[width=0.98\textwidth]{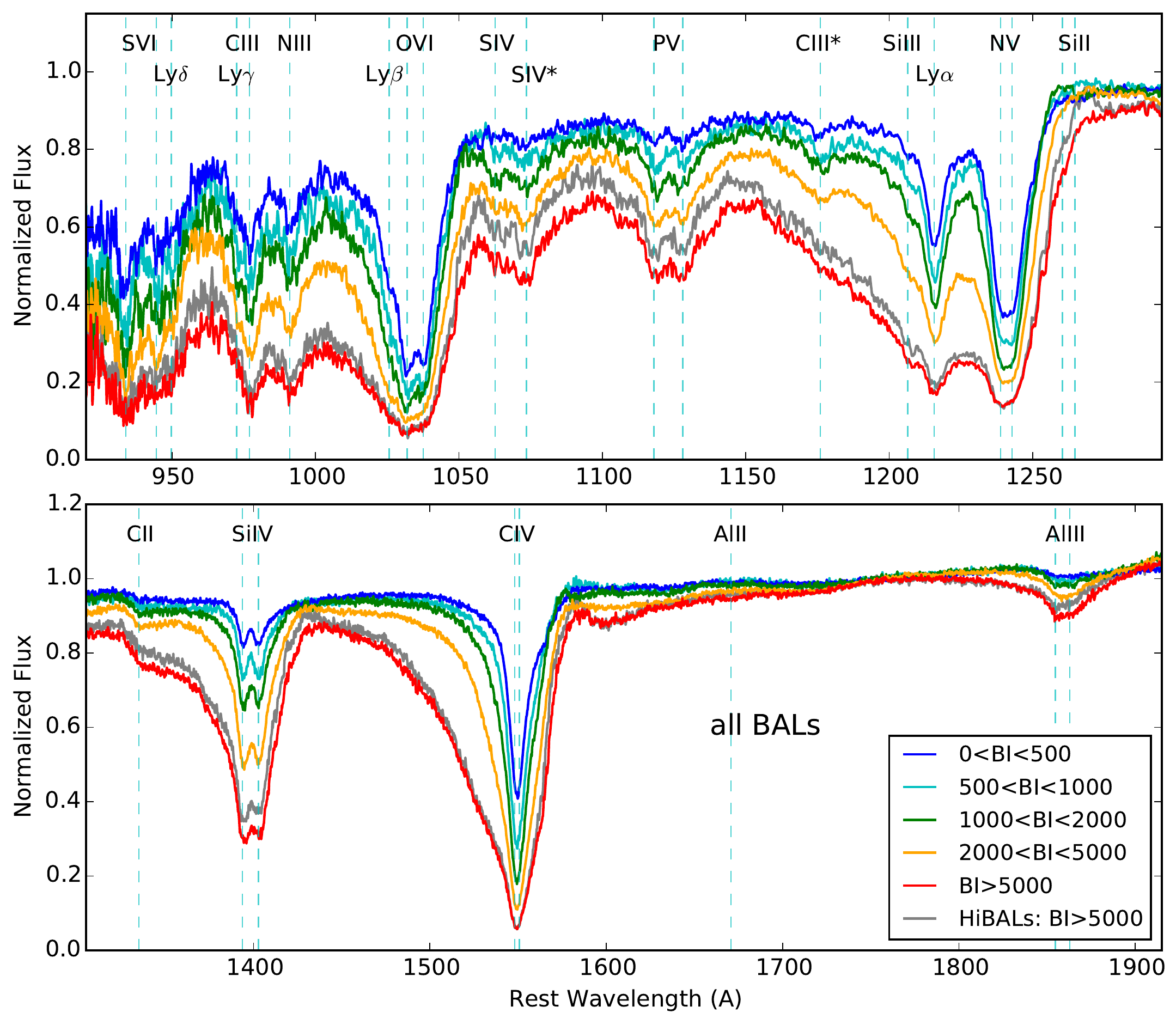}
\vspace{-7pt}
 \caption{Absorber-frame composite spectra of the `all BAL' subsamples sorted by \civ\ BAL strength (BI). The grey curve shows additionally the HiBAL composite with BI $>$ 5000 (see Table 1 and Section 2.1). Spectra in the top panel are smoothed slightly with a Gaussian filter to improve clarity. The dashed vertical lines mark the wavelengths of absorption lines in the rest frame define by \civ . Notice the resolved doublets with $\sim$1:1 depth ratios in \ovi\ \lam 1032,1038, \pv\ \lam 1118,1128, and \siiv\ \lam 1394,1403 and the excited-state lines \siiv * \lam 1073 and \ciii * \lam 1076.}
\end{figure*}

\begin{figure*}
\includegraphics[width=0.98\textwidth]{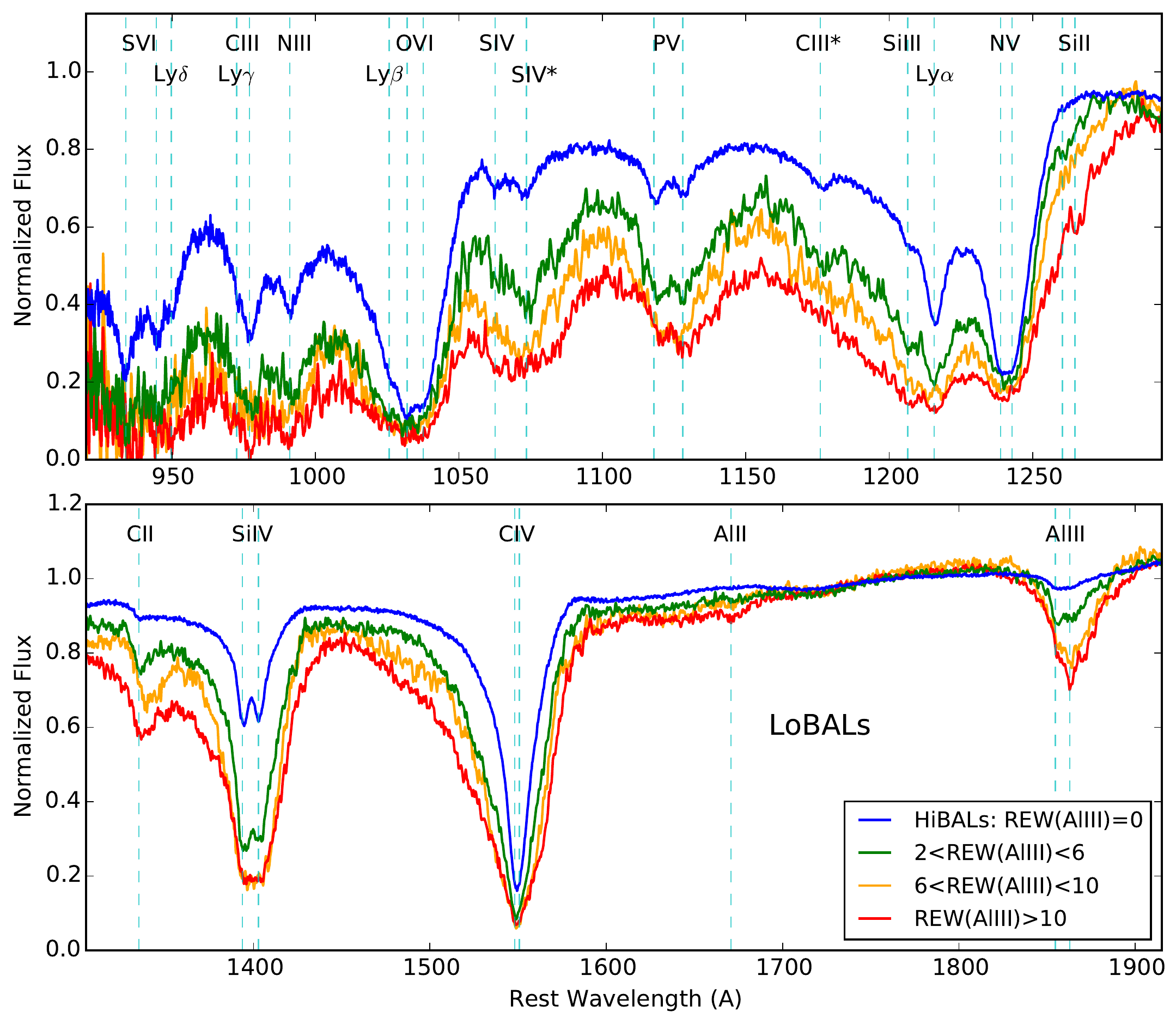}
\vspace{-7pt}
 \caption{Absorber-frame composite spectra of LoBAL quasars (detected in \aliii ) compared to a HiBAL sample with REW(\aliii ) = 0. See Figures 2 and 4 for additional notes.}
\end{figure*}

\begin{figure*}
\includegraphics[width=0.98\textwidth]{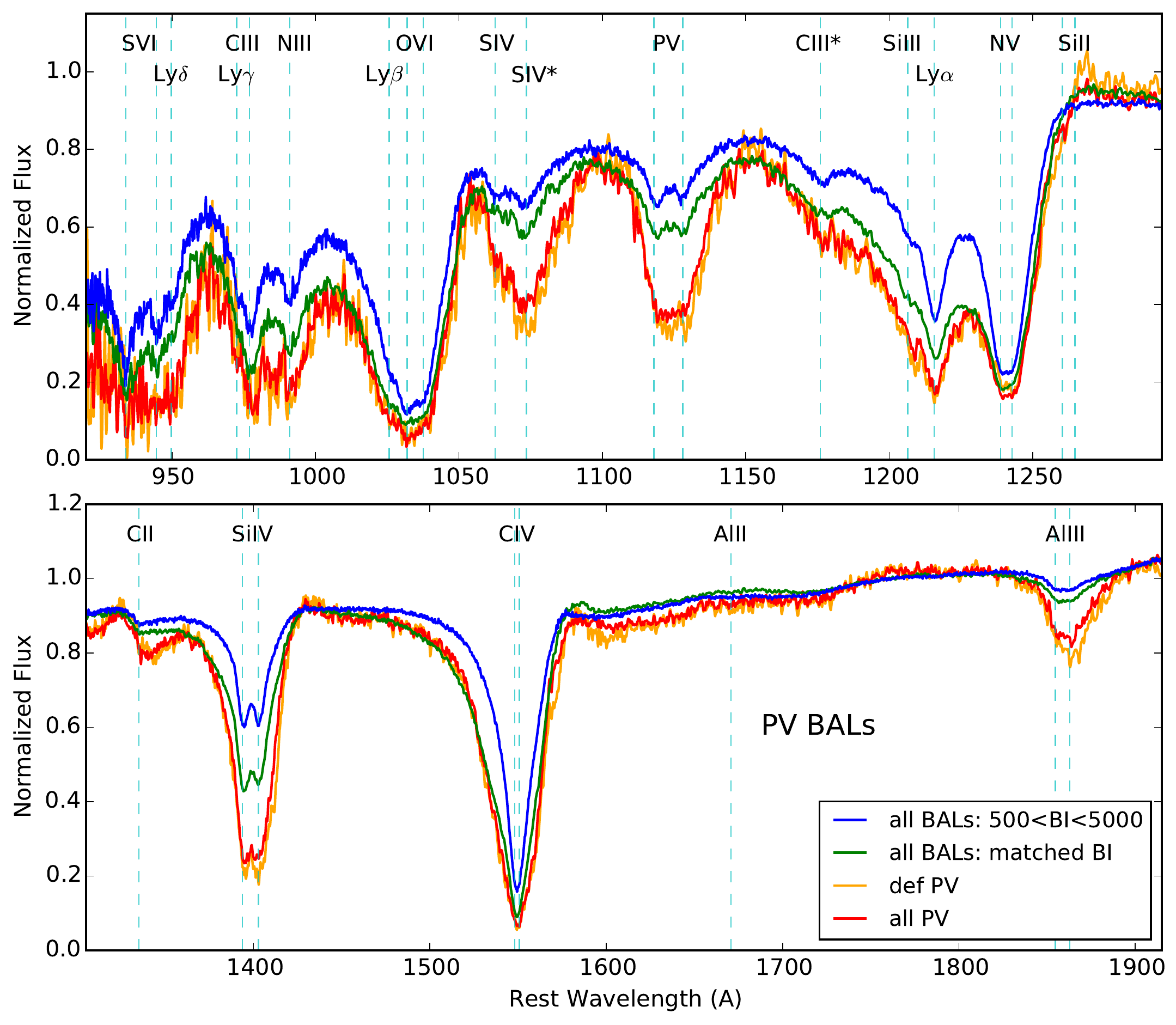}
\vspace{-7pt}
 \caption{Absorber-frame composite spectra of \pv -selected BAL quasars compared to an `all BAL' composite with 500~$>$~BI~$>$~5000 and another that is matched in BI to the \pv -selected BAL quasars. See Figures 2 and 4 for additional notes.}
\end{figure*}

\begin{figure*}
\includegraphics[width=1.0\textwidth]{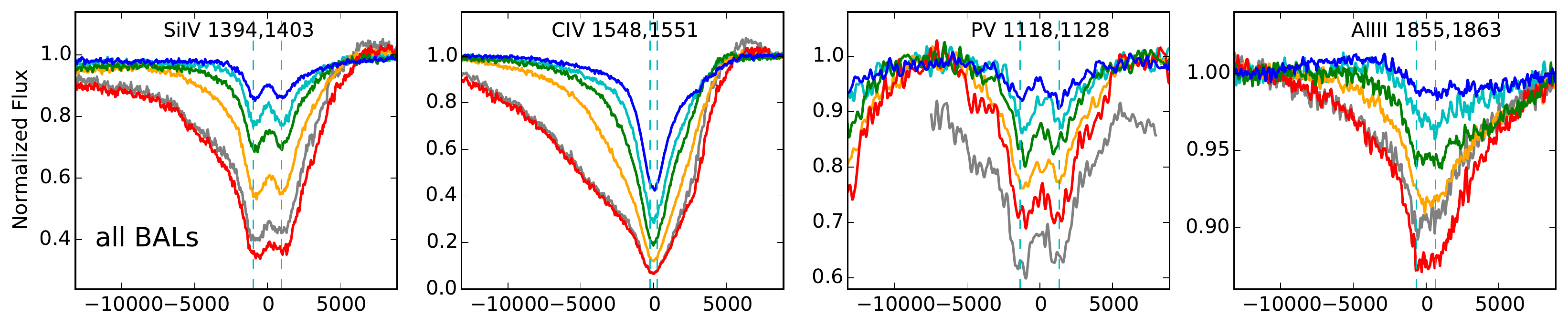} 
\includegraphics[width=1.0\textwidth]{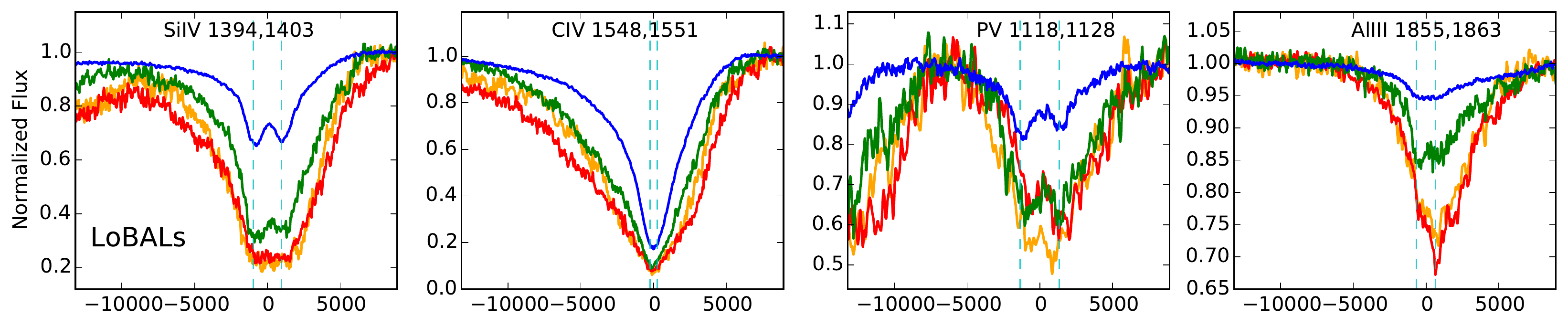} 
\includegraphics[width=1.0\textwidth]{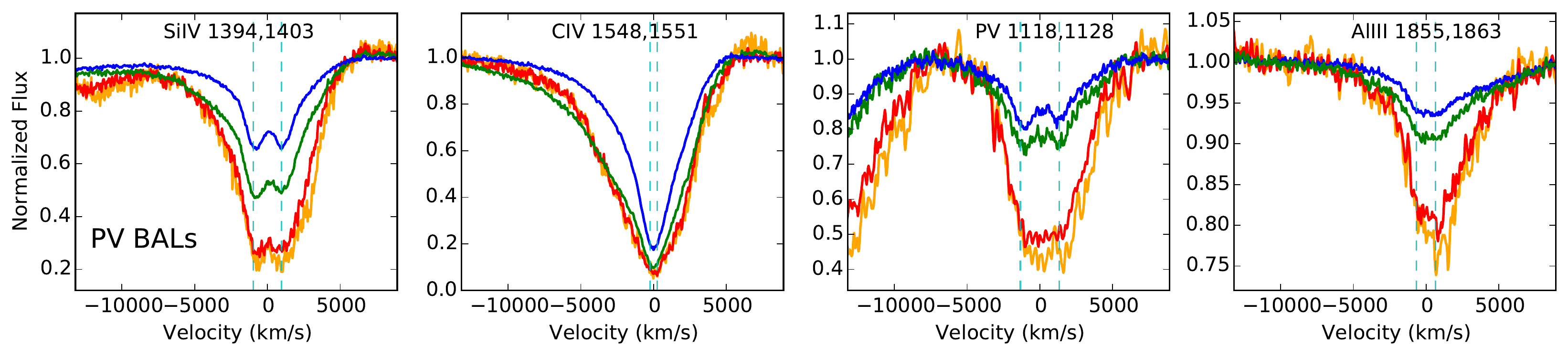} 
\vspace{-15pt}
 \caption{Absorber-frame BAL profiles normalized to the local continuum for the composites shown in Figure 4 to 6 (using the same color codes for different spectra). \textit{Top row}: all BALs (as in Figure 4). The \pv\ profile for the HiBALs with BI~$>$~5000 (grey curve) is shifted down by 0.12 in normalized flux to avoid overlap with the other composites. \textit{Middle row}: LoBALs vs. HiBALs (Figure 5). \textit{Bottom row}: \pv\ BALs vs. all BALs (Figure 6). The velocities are relative to the doublet midpoints in the absorber frame defined by the maximum depth in \civ\  (Section 3). The blue dashed vertical lines mark the separate doublet positions in that frame. Some of the spectra are smoothed with a Gaussian filter for clarity.}
\end{figure*}

Table 2 lists measured properties of the normalized line profiles plotted in Figure 7. The full widths at half minimum, FWHMs, in Table 2 represent the measured profile widths \textit{minus} the velocity separation of the doublet, e.g., from 498 \kms\ for \civ\ to 2677 \kms\ for \pv. These modified FWHMs provide consistent measures of the outflow velocity dispersions in the different lines. However, it is important to keep in mind that the velocity widths in these median spectra can have contributions from both the intrinsic profile widths and object-to-object differences in the line shifts relative to the absorber frame defined by \civ . The effects of different velocity shifts are most notable in the median \aliii\ profiles, which tend to have a broad red wing and (in LoBALs and large BI samples) a centroid offset toward longer wavelengths (Figure 7). These offsets are consistent with previous studies showing that low-ionisation BALs like \mgii\ tend to have smaller velocity shifts than higher ions like \civ\ \citep{Voit93, Hall02, Hall03}. 

\begin{table*}
 \centering
  \caption{Measured data for the \siv , \civ , \pv , and \aliii\ absorption lines in the absorber-frame composite spectra. Listed are the Sample name (as in Table~1), rest equivalent width, REW, a modified full width at half minimum, FWHM (after subtracting the doublet separations), and maximum line depth in the normalized spectrum. The $1\sigma$ uncertainties (not including continuum placement) on all three measured parameters are $\lesssim$5\% for lines with REW~$>$~10 \AA , $\lesssim$10\% for 1~$<$~REW~$<$~10~\AA , and $\lesssim$20\% for REW~$<$~1 \AA . Numbers for \pv\ in parentheses are lower limits due to continuum underestimates; no \pv\ data are listed for the strongest LoBALs due to blending problems (Section 4.2). }
  \begin{tabular}{r c c c c c c c c c c c c}
  \hline
    & \multicolumn{3}{c}{----------- \siiv\  ------------}& \multicolumn{3}{c}{------------ \civ\  ------------}& \multicolumn{3}{c}{------------ \pv\ ------------}& \multicolumn{3}{c}{------------ \aliii\ -----------}\\
  & REW & FWHM & depth & REW & FWHM & depth & REW & FWHM & depth & REW & FWHM & depth\\
   Sample & (\AA) & (km/s) &  & (\AA) & (km/s) & & (\AA) & (km/s) & & (\AA) & (km/s) & \\
 \hline
\underbar{mini-BALs:} & 2.4 & 1394 & 0.16 & 8.5 & 1300 & 0.70 & 1.1 & 1460 & 0.08 & 0.8 & 1733 & 0.02\\
 \hline
 \underbar{all BALs:}\\
0$<$BI$<$500 & 4.0 & 2254 & 0.14 & 10.9 & 1976 & 0.56 & 1.7 & 2587 & 0.09 & 0.4 & 4558 & 0.01\\
500$<$BI$<$1000 & 5.9 & 2073 & 0.22 & 14.9 & 2472 & 0.70 & 2.4 & 1867 & 0.13 & 1.1 & 3542 & 0.04\\
1000$<$BI$<$2000 & 8.6 & 2432 & 0.30 & 19.1 & 3100 & 0.80 & 3.5 & 1871 & 0.19 & 2.4 & 5346 & 0.06\\
2000$<$BI$<$5000 & 13.5 & 2882 & 0.45 & 27.3 & 4716 & 0.87 & (5.0) & (2541) & (0.23) & 3.5 & 3748 & 0.08\\
BI$>$5000 &  25.0 & 4809 & 0.65 & 45.0 & 8236 & 0.93 & (6.7) & (2629) & (0.29) & 5.6 & 4328 & 0.12\vspace{-5pt}\\ 
\multicolumn{13}{r}{.............................................................................................................................................................................................................................}\\
500$<$BI$<$5000 & 9.3 & 2437 & 0.34 & 20.2 & 3280 & 0.82 & 3.9 & 2091 & 0.20 & 2.6 & 4615 & 0.07\\ 
\hline
\underbar{LoBALs:} \\
REW(\aliii )=0 & 9.9 & 2431 & 0.34 & 23.0 & 3369 & 0.82 & 3.8 & 2134 & 0.19 & 2.3 & 3749 & 0.05\\
2$<$REW(\aliii )$<$6& 23.3 & 3922 & 0.68 & 35.1 & 6148 & 0.90 & 8.8 & 3694 & 0.38 & 5.2 & 2710 & 0.15\\
6$<$REW(\aliii )$<$10& 28.7 & 5132 & 0.79 & 41.4 & 6957 & 0.93 & (10.9) & (3428) & (0.41) & 8.3 & 3162 & 0.27\\
REW(\aliii )$>$10& 31.5 & 6081 & 0.77 & 47.8 & 8784 & 0.92 & --- & --- & --- & 9.4 & 2839 & 0.30\vspace{-5pt}\\ 
\multicolumn{13}{r}{.............................................................................................................................................................................................................................}\\
6$<$REW(\aliii )$<$24& 30.2 & 5404 & 0.76 & 43.8 & 7581 & 0.92 & (10.4) & (3312) & (0.43) & 8.2 & 2937 & 0.28\\
\hline
\underbar{HiBALs:}\\ 
BI$>$5000 & 21.8 & 4265 & 0.60 & 42.3 & 7787 & 0.93 & (6.0) & (2583) & (0.27) & 4.5 & 3736 & 0.10\\
\hline
\underbar{\pv -selected BALs:}\\
all BALs: matched& 15.9 & 3240 & 0.52 & 32.0 & 5609 & 0.90 & 5.7 & 2630 & 0.25 & 4.0 & 3746 & 0.09\\
definite \pv & 24.3 & 3967 & 0.78 & 32.6 & 5970 & 0.92 & 12.9 & 3316 & 0.58 & 7.5 & 3368 & 0.23\\
all \pv & 21.9 & 3559 & 0.73 & 32.0 & 5567 & 0.92 & 11.6 & 3087 & 0.51 & 6.9 & 3382 & 0.21\\
\hline
\end{tabular}
\end{table*}

Figure 8 shows normalized absorber-frame composite spectra for the mini-BALs (Section 2.2), the large moderate BAL sample defined by 500~$>$~BI~$>$~5000 (Section 2.1), and the moderate-to-strong LoBALs with $6<\textrm{REW(\aliii)}<24$ \AA\ (Section 2.3). This plot helps to show the progression in absorption line strengths across the outflow types from mini-BALs to BALs to LoBALs. The normalization is achieved using smooth curves constrained to match the median flux in narrow wavelength windows that appear free of absorption. The true quasar continua in the BAL spectra are poorly determined at short wavelengths due to overlap between adjacent BALs (see Figures 4 and 5). We manually place the continuum above the measured spectrum at these wavelengths, $\lesssim$1100 \AA , but still conservatively low to produce conservatively weak depictions of the absorption lines in Figure 8.

\begin{figure*}
\includegraphics[width=1.0\textwidth]{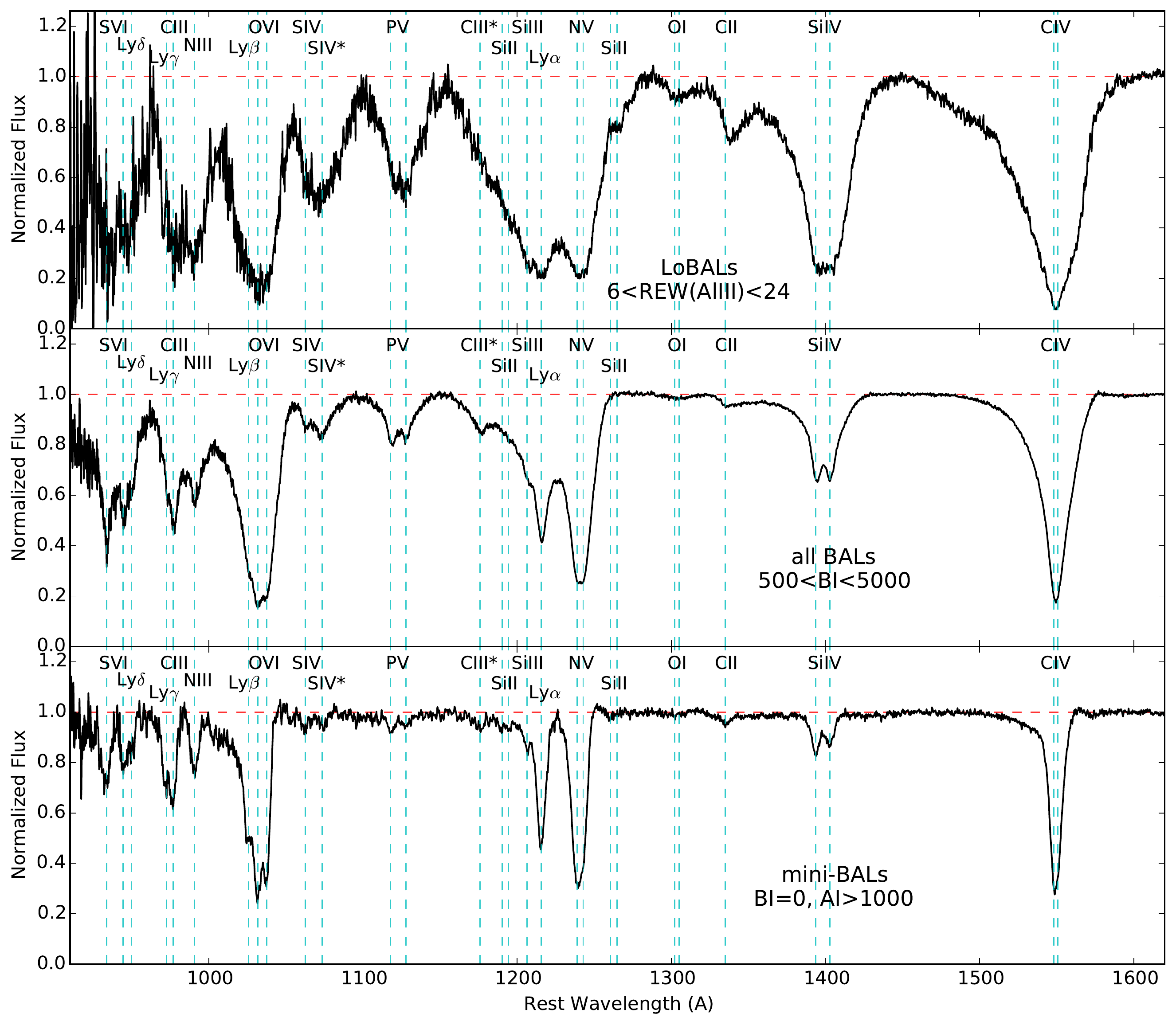} 
\vspace{-18pt}
 \caption{Normalized absorber-frame composite spectra of three outflow subsamples. \textit{Top panel}: LoBALs with $6<\textrm{REW(\aliii)}<24$ \AA\ (Section 2.3). \textit{Middle panel}: All BALs with intermediate strengths defined by $500 <$~BI~$< 5000$ (Section 2.1). \textit{Bottom panel}: Mini-BALs defined by BI~=~0, AI~$> 1000$, and \texttt{vmin\_450}~$>1000$ \kms\ (Section 2.2).}
\end{figure*}

The absorber-frame composite spectra in Figures 4 to 8 reveal numerous outflow lines that are difficult or impossible to measure in individual BOSS spectra, e.g., at wavelengths $<$1216 \AA\ in the \lya\ forest. This includes important diagnostic lines like the low-abundance \pv\ \lam 1118,1128 doublet that places firm lower bounds on the outflow column densities, and the excited-state lines \siv * \lam 1073 and \ciii * \lam 1176 that constrain the outflow densities and locations. Also, remarkably, the composite spectra have resolved doublet components with roughly 1:1 strength (depth) ratios in \pv\ \lam 1118,1128, \siiv\ \lam 1394,1403, and (in some quasars) \ovi\ \lam 1032,1038. We discuss the physical implications for quasar outflows in Section 5. Here we describe the main empirical results.

\subsubsection{BALs \& Mini-BALs: trends with outflow line strength}

The mini-BALs in Figure 8 and `all BAL' composites in Figures 4 and 7 show the progression of outflow line properties across the full range of \civ\ strengths in our study from BI~=~0 to BI~$>$~5000. The median line profiles become generally broader and deeper with increasing BI, as expected. However, it is also interesting that \siiv\ and especially \civ\ have increasingly stronger blueshifted wings, e.g., in composites with BI $>$ 2000. The strengths of broad blue wings at large BI depend on the line ionisation and/or optical depth. In particular, the low-abundance and lower-optical depth \pv\  \lam 1118,1128 lines are roughly symmetric while broad blue wings are present in \siiv , stronger \civ, and stronger still in higher-ion lines \nv\ \lam 1239,1243 and \ovi\ \lam 1032,1038. (The blue wings in \nv\ and \ovi\ are both blended with other lines [Figure 4], but the gradual recovery in the flux blueward for these lines suggests that it is caused primarily by broad blueshifted wings rather than unrelated blends.) 

Similarly, the median line depths also depend on ionisation and optical depth. In particular, \pv\ and \siiv\ become much deeper with increasing BI while \nv\ and especially \ovi\ are deep in all of the composites with little variation, while \civ\ is intermediate. Table 2 shows that the \aliii\ line depth increases by a factor of $\sim$10 across the range from mini-BALs/weak BALs to strong BALs, the \siiv\ and \pv\ depths increase by factors of $\sim$4, and \civ\ increases by less than a factor of two. The \nv\ and \ovi\ lines have even less depth variation because they are strong in all of the composites. 

The mini-BALs (Figure 8) are not a straightforward extension of these trends from strong to weak BALs. They are distinctly narrower than all of the BALs (Table 2), as expected, but they also have deeper absorption in \civ, \nv, and \ovi\ than the weak BAL composites. This might indicate that the mini-BAL outflows have characteristically different physical conditions, but they might also be tied to selection effects in the mini-BAL subsample. In particular, the defining requirements for BI = 0 and AI $>$ 1000 combine to select narrow and \textit{deep} \civ\ troughs that will naturally be accompanied by narrow and deep \nv\ and \ovi\ lines. It is also possible that our mini-BAL subsample is contaminated by unrelated intervening absorption lines, which would also favor deep absorption in \civ , \nv, and \ovi, weak \siiv , and no \pv\ \citep[][and Perrotta et al., MNRAS, in press]{Perrotta16}.

It is important to emphasize that the low-abundance \pv\ lines are detected in all of our outflow composite spectra. This includes weak BALs and mini-BALs (see Figures 4 to 8). \pv\ outflow lines are rarely measured in individual quasars due to the short wavelength and blending problems in the \lya\ forest. Our results show that this line is common, with a median observed depth that is roughly half to two thirds of the \siiv\ line. This assessment of the relative strength is consistent with \pv\ measurements in individual outflow quasars \citep[e.g.,][]{Hamann98, Borguet12, Chamberlain15, Capellupo17, Moravec17}. Our composite spectra also show that the \pv\ and \siiv\ doublets, when resolved, typically have observed depth ratios near 1:1 regardless of the absolute line depths or the outflow line type based on BI and AI. We also note that while the \pv\ absorption in our mini-BAL composite appears at a relatively low velocity of $\sim$2000 \kms , \cite{Hamann18} showed recently that \pv\ absorption with a $\sim$1:1 depth ratio is also present in a median composite spectrum of weak BAL/mini-BALs at exclusive large velocity shifts $\gtrsim$6000 \kms . 

\subsubsection{LoBALs versus HiBALs}

The absorber-frame composite spectra of LoBALs shown in Figures 5 and 8 differ from the `all BAL' and HiBAL subsamples in several important ways: 

1) LoBALs typically have very strong \civ\ and \ovi\ BALs. In particular, all of the LoBAL subsamples in our study have median BI~$>$~4400, which is more than twice the value BI~=~2118 for HiBAL quasars selected the same way (but with REW(\aliii ) = 0, see Table 1). The \ovi\ strengths are difficult to quantify due to blends, but this line is clearly deep and strong in LoBALs compared to HiBALs (c.f., Figures 4 and 5). 

2) LoBALs have stronger absorption in all low-ionisation lines including \cii\ \lam 1335, \alii\ \lam 1670, and \feii\ \lam 2344,2383,2600, and \mgii\ \lam 2796,2804 (not shown), in addition to stronger \aliii . This confirms our assertion in Section 2.3 that \aliii\ is a reasonable indicator of LoBAL quasars. 

3) At a given BI value (\civ\ BAL strength), LoBALs have stronger absorption in lines of intermediate-ionisation and low-abundance ions like \siiv\ and \pv . The feature we identify as \pv\ absorption in the strongest LoBALs with REW(\aliii)~$>$~10 \AA\ might have significant contributions from a series of lines in the ground multiplet of \feiii\ at 1122.5 \AA\ to 1130.4 \AA . However, a careful overlay of those lines on top of the observed line trough (not shown), together with consideration of the transition strengths, indicates that most of the absorption at these wavelengths is due to the \pv\ doublet. 

4) Low-ionisation outflow lines in all of the BAL composites appear redshifted with respect to the absorber frame defined by the \civ\ flux minimum (Figures 5 to 7). In well-measured/strong lines of \cii , \aliii , \feii , and \mgii, the median redshifts relative to \civ\ are in the range $\sim$800 to $\sim$2200 \kms . These low-ion line profiles are also asymmetric, with extended red wings, indicating that substantial fractions of the individual quasars have even larger redshifts. These results are consistent with previous studies showing that \mgii\ BALs tend to concentrate at lower velocities than the \civ\  BALs measured in the same spectra \citep[e.g.,][]{Voit93}. 

5) The LoBAL composite spectra have weaker fluxes than the all BAL and HiBAL composites at all wavelengths blueward of $\sim$1240 \AA . We attribute this to the LoBALs having both redder continuum slopes at the quasars (Section 4.1.1) plus broader high-ionisation BALs with numerous weak BALs that combine to produce overlapping absorption troughs in the near-UV. 


\subsubsection{PV-selected BALs} 

The \pv-selected BAL quasars from \citep{Capellupo17} are a valuable subsample because they emphasize the large outflow column densities needed for \pv\ absorption. \cite{Capellupo17} noted that these quasars typically have strong \pv\ absorption accompanied by strong \civ\ due to a selection effect for securely identifying \pv\ absorption in the \lya\ forest (Section 2.4). Our absorber-frame composite spectra show further that their line profiles and line strength ratios are also notably different than other BAL quasars. In particular, compared to the `all BAL' composite with similar BI (Figures 6 and 7), the \pv-selected quasars have stronger (deeper) absorption in various low-ionisation or low-abundance lines such as \cii , \mgii , \aliii , \siv, \siv *, and \siiv . 

Surprisingly, the \aliii\ troughs in the \pv-selected composites are nearly as strong as in the strongest LoBALs, which are \textit{defined} to have strong \aliii\ absorption (Figures 5 to 7, also Table 2). In particular, the \aliii\ troughs in the \pv-selected samples have measured depths roughly 75 per cent of \aliii\ in the strongest LoBAL composites (Figure 7). In this sense, the \pv-selected BALs are intermediate between HiBALs and LoBALs but more similar to strong LoBALs then they are to HiBALs.  It is also notable that the strong \pv\ lines in the \pv-selected quasars, along with \aliii\ and other low-ionization BALs, appear typically redshifted with extended red wings in the absorber frame defined by \civ. These shifts were also noted above for the LoBAL composites (Section 4.2.2). These empirical relationships between LoBALs and \pv-selected BALs indicate that the large column densities required for \pv\ absorption are an important factor in the LoBAL outflow phenomenon. We will return to this point in Section 5.2. 


\section{Implications for Quasar Outflows}

The median composite spectra presented in Section 4 place important constraints on the outflow physical properties and the relationships between the different observational classes of mini-BALs, HiBALs, and LoBALs. Here discuss the main results.  

\subsection{Quasar SEDs \& radiative acceleration}

The trend described in Section 4.1.3 for stronger and broader BALs in quasars with weaker high-ionisation emission lines, notably \heii\ \lam 1640 and \civ\ \lam 1549,1551, can be attributed to the relationship of both features to the UV continuum shape \citep{Baskin13,Baskin15}. It is known generally from photoionisation studies have that harder UV spectra (with relatively more far-UV flux) produce larger rest equivalent widths in high-ionisation near-UV emission lines \citep{Netzer92, Zheng95, Korista97}. The \heii\ line is a particularly good indicator of UV spectral slope because it is produced directly by recombination and, therefore, its REW is directly tied the flux in \heii-ionising photons at $>$54 eV relative to the near-UV continuum flux at 1640 \AA\ (7.56 eV). The connection to quasar outflows depends more subtlely on ionisation effects in radiatively-accelerated winds. In particular, hard UV spectra produce higher degrees of ionisation in the outflows and therefore lower opacities for a given amount of near-UV flux available to drive the flow. The result is less effective radiative driving \citep{Murray95, Chelouche03, Leighly04b, Proga04, Baskin13}. The ideal spectrum for radiative acceleration has a soft UV slope: strong in near-UV where the resonance absorption lines are located but relatively weak in the far-UV to avoid overionisation \citep[see also][]{Laor14}. \cite{Baskin13} and \cite{Baskin15} showed previously that weak \heii\ emission-line REWs (and thus softer far-UV spectra) are strongly correlated with larger BAL outflow speeds. Other studies have shown, similarly, that weak \heii\ and \civ\  emission-line REWs also correlate with larger \civ\  emission-line blueshifts \citep[another indicator of quasar outflows, albeit in the broad emission-line regions,][]{Leighly04b, Leighly07b, Richards11, Wu12, Coatman16, Sun18}. 

Our study extends this work to show that LoBAL quasars are at an extreme in these trends. Their median spectra have the broadest quasar-frame BAL troughs and the highest outflow velocities accompanied by the weakest \civ\ emission lines, no measurable \heii\ emission, and \textit{enhanced} \feii\ emission. Thus they support the claims that soft far-UV continua (weak \heii\ emission) are the main driver for high-speed quasar outflows. However, the LoBALs in our study do not follow another trend identified previously by \cite{Baskin15} for larger BAL velocity shifts in quasars with \textit{bluer} near-UV spectra (e.g., at wavelengths $\sim$1700 \AA\ to $\sim$3000 \AA). Baskin et al. attribute this to viewing angle effects, where higher outflow speeds are attained at higher latitudes above the accretion disk plane (subject to less dust reddening along our lines of sight). We find that LoBAL quasars exhibit the opposite behavior; they have the largest median velocity shifts and the \textit{reddest} near-UV spectra (Figures 2 and 3). This does not necessarily conflict with the results in \cite{Baskin15} because they examine only HiBAL quasars with LoBALs excluded. However, LoBALs require a different interpretation. They might identify situations where the soft UV continua that cause weak \heii\ emission and high outflow speeds occur along dust-reddened sightlines \textit{near} the accretion disk plane. 

The physical explanation for softer UV spectra in outflow quasars is not known. It might occur in quasars with normal/hard far-UV spectra if their emitted far-UV flux is filtered through large column densities of highly-ionised gas at the base of the outflows \citep{Murray95, Proga04, Proga07, Sim10}. Large amounts of this `failed wind' material could be present if it is too ionised and too transparent to be radiatively driven, but it might serve as a radiative shield to produce the spectral softening needed for radiative acceleration of the outflow material downstream. However, one recent study indicates that quasar outflows can be accelerated to high speeds without substantial radiative shielding and, moreover, that the shielding gas should produce strong near-UV absorption lines at v$\,\sim 0$ that are not observed \citep{Hamann13}. Therefore, \textit{intrinsically} softer far-UV spectra emerging from the accretion disks should also be considered \citep[e.g.,][]{Laor14}. If quasars naturally have a range of intrinsic/emitted UV spectral shapes, then outflows with higher speeds and larger column densities could naturally occur more often in objects with softer UV continua \citep[see also][]{Richards11}. 

Finally, we note that the relationship of the emission lines to outflow properties cannot be explained simply by luminosity effects because all of the quasar subsamples in our study have similar median absolute magnitudes ($M_i$, Table 1). This rules out any direct connection to the well-known Baldwin effect, which describes a trend for smaller emission-line REWs in more luminous quasars \citep{Baldwin77, Kinney90, Dietrich02, Shields07}. However, the Baldwin effect has also been attributed to characteristically softer far-UV spectra in more luminous quasars \citep[e.g.,][]{Korista98}. Therefore, the diverse emission-line strengths correlated with BAL properties at fixed $M_i$ in our study might be related to \textit{scatter} in the Baldwin effect caused in by scatter in the intrinsic/emitted UV spectral shapes.

\subsection{Total column densities \& ionisation parameters}

One of the main results of our study is that \pv\ \lam 1118,1128 absorption is present in all of the median outflow composites from mini-BALs to the strongest BALs (Figure 4 to 8). It is also remarkable that the \pv\ doublets are resolved in most of our absorber-frame composite spectra with depth ratios near $\sim$1:1. This implies that the \pv\ lines are \textit{typically} optically thick. Only the two strongest LoBAL composites are ambiguous in this regard because their BALs broad and blended so that the \pv\ doublet is not resolved. However, these LoBAL composites have stronger \pv\ absorption that other BAL composites where $\sim$1:1 ratios are clearly measured. We conclude that strong LoBALs are also very likely to have saturated \pv\ absorption. 

These results require large outflow column densities in an intense ionizing radiation field. If we adopt a minimum optical depth of $\tau(1128) \gtrsim 3$ in the weaker \pv\ \lam 1128 line and a velocity dispersion in the \pv-absorbing gas given by FWHM~$\sim 2200$ \kms\ (Table 2), then the photoionisation calculations in \cite[][and Hamann et al., submitted, based on solar abundances and standard quasar ionising spectra]{Leighly11} indicate that BAL outflows typically have \textit{minimum} total column densities $\log N_H (\textrm{cm}^{-2}) \gtrsim 22.7$ and \textit{minimum} ionisation parameters $\log U \gtrsim -0.5$. The ionisation parameter is defined as the dimensionless ratio of the hydrogen-ionising photon density to total hydrogen gas density (\hi\ + \hii ) at the illuminated face of the outflow clouds \citep[e.g.,][]{Ferland13}. Very similar $N_H$ and $U$ values are inferred from so-called `radiation pressure confinement' models of BAL outflows, which allow the densities and small-scale spatial structure to adjust to the local pressure balance \citep{Baskin14}. Large ionisation parameters are essential in all cases to fully ionise large column densities of gas to produce the observed \pv\ absorption. Mini-BALs are narrower than BALs so they can reach the same line optical depths at lower column densities. However, the difference in FWHM is less than a factor two (Table 2), such that the typical mini-BAL in our study also requires large \textit{minimum} total column densities $\log N_H (\textrm{cm}^{-2}) \gtrsim 22.5$. These lower limits on $U$ and $N_H$ are consistent with previous estimates for individual quasars with \pv\  BALs \citep[e.g.,][although the Chamberlain et al. study appears to underestimate the column densities by underestimating the saturation effects]{Chamberlain15, Capellupo14, Capellupo17, Moravec17}. The main new result from our study is that saturated \pv\ absorption and large $U$ and $N_H$ values are \textit{common} in all outflow classes from mini-BALs to the strongest BALs and LoBALs. 

\subsection{Partial covering and spatial structure}

Saturated absorption in the low-abundance \pv\ \lam 1118,1128 lines also indicates that all of the other common outflow lines like \civ , \siv , \nv , and \ovi\ are extremely optically thick \citep{Hamann98, Leighly11, Borguet12}. In particular, our result for $\tau (1128) \gtrsim  3$ based on $\sim$1:1 doublet ratios in \pv\ (Section 5.2) indicates that the weaker component \lam 1551 in the \civ\ doublet has a minimum optical depth of $\tau(1551) \gtrsim 3000$ if the C/P abundance ratio is roughly solar (see refs above, also Hamann et al. submitted). These predictions based on photoionisation models are supported directly by our absorber-frame composite spectra showing $\sim$1:1 depth ratios in other widely-separated doublets of \siiv\ and \ovi\ (Figures 4 to 8). The \aliii\ \lam 1855,1863 doublet has a smaller velocity separation than \pv\ and \siiv\ but it also appears to have depth ratios consistent with $\sim$1:1 in most of our weaker/narrow BAL composites. We also find that the Lyman series lines, when resolved, also have depth ratios indicative of saturation. See for example the normalized mini-BAL and BAL composite spectra shown in Figure 8. The mini-BAL composite in this figure has observed depth ratios in  \lya~:~\lyb~:~\lyg~:~\lyd\ of roughly $1.0:0.8:0.7:0.3$ (not correcting for possible blends), which are clearly not compatible with the optically thin ratios of $1.0:0.16:0.06:0.03$ \citep{Wiese09}. 

 
We conclude that all of the strong outflow lines are optically thick, but none of them reach zero intensity in our median composite spectra. They can actually be quite weak, such as the \siv\ and \pv\ doublets in the mini-BAL and weak BAL (small BI) composites  (Figures 7 and 8). This is a signature of partial covering of the background light source where unabsorbed continuum flux fills in the observed absorption troughs \citep[e.g.][]{Hamann97, Arav99, Arav99b, Gabel03}. For optically thick lines, their observed depths below the continuum provide direct measures of the line-of-sight covering fractions, $C_0$ (where $0< C_0\leq 1$). Most of the outflow lines absorb the quasar continuum away from the broad emission lines (Figures 2 and 3). Therefore, the emission source partially covered by the outflow gas is the UV-emitting accretion disk whose diameter is of order 0.006 pc for quasars with luminosities like our samples (see Hamann et al., submitted, and refs. therein).  This places an upper limit on the characteristic transverse size of absorbing structures in the outflows. 

Another important result is that different optically-thick lines in the absorber-frame composite spectra reach different observed depths indicative of different covering fractions. For example, compare the normalized depths of the \pv , \siiv, and \civ\ doublets in Figures 7 and 8 and Table 2. The \civ\ covering fractions increase steadily across the `all BAL' subsamples from $C_0\approx 0.56$ in the weakest BALs with 0 $<$ BI $<$ 500 to $C_0\approx 0.93$ in the strongest BALs with BI $>$ 5000. The \siiv\ covering fractions are smaller but change more dramatically than \civ , ranging from 0.14 to 0.65 in the weakest to strongest BALs, respectively. \pv\  roughly tracks the \siiv\ behavior with line depths and covering fractions that are typically $\sim$1/2 to $\sim$2/3 of the \siiv\ values. The modified FWHMs in Table 2 show further that the weaker \siiv\ and \pv\ lines form in regions with lower velocity dispersion than \civ , at least in the moderate to strong BAL composites. The mini-BALs in our study also exhibit a very wide range of optically-thick line depths/covering fractions from 0.70 in \civ\ to only $\sim$0.08 in \pv. 

These results identify partial covering by a clumpy inhomogeneous medium that presents a range of column densities and optical depths across the projected area of the continuum source \citep{deKool02, Hamann01, Hamann04, Arav05, Arav08}. Schematic illustrations of this type of absorbing geometry can be found in \cite{Hamann01} and \cite{Hamann04}. The main idea is that stronger transitions in abundant ions like \civ\ \lam 1549,1551 have larger covering fractions and therefore deeper observed troughs because they are optically thick over larger projected areas. There can be optical depth-dependent line depths and profiles in observed spectra even if all of the lines are optically thick. Among the lines we measure in a given spectrum, \pv\ traces only high-column density gas that resides in smaller clumps/regions covering smaller spatial areas (to produce shallow absorption troughs) and smaller ranges in velocity (to produce narrower line profiles). Strong lines like \civ\ trace also lower-column density gas that is more dispersed in space and velocity to produce observed lines with deeper troughs with broader profiles. The characteristic size of the absorbing clumps must be $\lesssim$0.006 pc for strong lines like \civ\ and several times smaller for \pv. \cite{Moravec17}, Hamann et al. (in prep.), and the other studies cited above have already invoked this type absorbing geometry to explain the complex behaviors of BALs and mini-BALs in individual quasar spectra. Our study based on composite spectra of diverse quasar subsamples shows that clumpy inhomogeneous absorbing geometries are \textit{common} in all types of BAL and mini-BAL outflows. 

\subsection{Densities \& radial distances}

Absorption lines from excited states provide valuable constraints on the outflow densities and distances from the quasars \citep{Hamann01, Moe09, Dunn10}. Most of the absorber-frame composite spectra in our study (Figures 4, 5, 6, and 8) exhibit \ciii * \lam 1175 absorption along with \siv * \lam 1073 at a strength that is generally stronger than its corresponding resonance line \siv\ \lam 1063. The only composites that do not show these lines are strong BALs and LoBALs where the detections are hampered by blending problems. Among the detections, only the mini-BAL composite has \siv * \lam 073 not clearly stronger than \siv\ \lam 1063 (Figure 8). The observed depth ratio in this case is roughly 1:1. 

 The \siv\ \lam 1063 and \siv * \lam 1073 lines arise from the ground multiplet at energies of 0 and 0.12 eV, respectively. If the lines are optically thin, the ratio of \siv * \lam 1073/\siv\ \lam 1063 absorption strengths will range from $\sim$0 in the low density limit to $\sim$1.9 at densities above $\sim$$10^6$ \cmn\ \citep{Leighly09, Chamberlain15}. These predictions have a weak temperature dependence that is negligible for our purposes. A more important factor is the line optical depths. We expect that at least the resonance line \siv\ \lam 1063 is saturated given that \pv\ absorption with a $\sim$1:1 doublet ratio is present in all of the composites where \siv\ and \siv * are detected. Saturation will push the observed \siv * \lam 1073/\siv\ \lam 1063 depth ratio closer to unity for any density (keeping in mind that very optically thick lines might never reach unity ratios in an inhomogeneous absorber, Section 5.3). 
 
 The observed \siv */\siv\ ratios in our median BAL composites lie in the range $\sim$1.5 to $\sim$2.0. This is near the high-density limit requiring conservatively $n_e \gtrsim 3\times 10^5$ cm$^{-3}$ regardless of saturation \citep[see Fig. 3 in][combined with transition data from the National Institute for Standards and Technology spectroscopic database\footnote{https://www.nist.gov/pml/atomic-spectra-database}]{Chamberlain15}. Saturation effects in this regime would cause us to underestimate the density, reinforcing the value above as a lower limit. In the mini-BAL composite (Figure 8), the \siv */\siv\ depth ratio is consistent with unity with an uncertainty of order 20 percent (due mainly to the uncertain continuum placement). This ratio requires a significant density to populate the excited state, but the specific constraint is uncertain. If we assume the line optical depths are not more than a few, then the minimum density is $n_e \gtrsim 10^4$ cm$^{-3}$

The \ciii * \lam 1175 line arises from a much higher energy metastable state $\sim$6.5 eV above ground. This level population controlled by collisional excitation has a strong temperature dependence. Detailed simulations of the absorption-line regions with possible saturation effects would be needed to disentangle the density and temperature dependence of the \ciii * \lam 1175 \AA\ line strength relative to other lines such as the resonant \ciii\ \lam 977 transition. However, \cite{gabel05} show that, for reasonable gas temperatures of order 10$^4$ K, significant populations in \ciii * excited state require densities above $n_e \gtrsim 10^5$ \cmn ,  consistent with the \siv */\siv\ results described above. 

The minimum density inferred from \siv*/\siv\ combined with the minimum ionisation parameter from \pv\ (Section 5.2) place a firm upper limit on the typical radial distance, $R$, of the outflows from the central quasars, namely,
\begin{equation}
R \ \lesssim \ 23\left({{\lambda L_\lambda (2100\textrm{\AA})}
\over{7.8\times 10^{45}\,\textrm{ergs~s}^{-1}}}\right)^{0.5}
\left({{3\times 10^5\,\textrm{cm}^{-3}}\over{n_H}}\right)^{0.5}
\left({{0.3}\over{U}}\right)^{0.5}~\textrm{pc}
\end{equation}
This expression is simply the definition of the ionisation parameter using a standard quasar ionising spectrum adopted from \citep[][see also Section 2.0 above]{Hamann13}. 

These density and distance results are in tension with some studies of individual quasars \citep[e.g.,][]{Dunn12, Arav13, Borguet13, Chamberlain15} and a recent survey by \cite{Arav18} suggesting that a majority of BAL quasars have line ratios \siv*/\siv\ $<$ 1 and, therefore, the outflows have lower densities and large distances $R>100$ pc than our estimates here. We do not explore the reasons for this discrepancy except to note that there are no selection biases in our study regarding the \siv* and \siv\ line strengths. Our samples include all well-measured BAL quasars listed in the BOSS DR12Q catalog based on their \civ\ absorption. Line ratios \siv*/\siv\ $>$ 1 indicating high densities are evident in all of our median composite BAL spectra not affected by severe blending. This includes the largest sample with $500<\textrm{BI}<5000$ that should be representative of most BALs (Figure 8). Only the mini-BALs out for having a poorly-measured median depth ratio \siv*/\siv~$\sim$~1, which leaves the interpretation of their typical densities and distances ambiguous. A minor factor contributing to our smaller derived distances compared to the other studies is that we adopt a larger ionisation parameter  required by our measurements of \pv\ absorption (Section 5.2).

\subsection{Energetics \& feedback}

Here we use the results above to estimate typical outflow kinetic energies for comparison to the requirements for feedback to galaxy evolution (Section 1). We approximate the global outflow geometry as an expanding shell with global covering factor $Q$ (i.e., the fraction of 4$\pi$ steradians covered by the outflow as seen from the central quasar). Then the minimum kinetic energies are 
\begin{equation}
K \ \gtrsim \ 4.8\times 10^{53} \; \left({{Q}\over{0.15}}\right)\left({{N_H}\over{5\times 
10^{22}\,\textrm{cm}^{-2}}}\right)\left({{R}\over{1\,\textrm{pc}}}\right)^2
\left({{\textrm{v}}\over{8000\, \textrm{km/s}}}\right)^2 ~~\textrm{ergs}
\end{equation}
where $N_H\gtrsim 5\times 10^{22}$ \cmN\ is a typical minimum total column density inferred from the saturated \pv\ troughs (Section 5.3) and $\textrm{v}=8000$ \kms\ is a typical outflow speed in strong BAL troughs (Section 4.1.2). $Q=0.15$ is an approximate global outflow covering factor based on the incidence of \civ\ BALs in SDSS quasars \citep[with estimates ranging from $\sim$0.15 to $\sim$0.41 depending on definitions and adjustments for the bias against BALs in SDSS,][]{Trump06, Knigge08, Gibson09, Allen11}. $R = 1$ pc is a placeholder radial distance that we adopt for illustration, but it is very roughly consistent with absorption-line variability studies \citep{Moe09, Hall11, Capellupo13, Capellupo14, Moravec17} and with theoretical models of radiatively-driven accretion-disk winds \citep{Murray95, Proga00, Proga04}. It is much smaller than the $>$0.1 to $>$1~kpc size scales derived from excited-state absorption lines in some other BAL outflow studies \citep[e.g.,][]{deKool01,Moe09, Dunn10, Arav13, Borguet13, Arav18}. 

We estimate the time-averaged kinetic energy output (i.e., the kinetic energy luminosity, $L_K$) by dividing $K$ by a characteristic flow time, $R/\textrm{v}\sim 122$ yr. This yields $L_K\gtrsim 1.2\times 10^{44}$ erg~s$^{-1}$ for the parameters adopted above. Dividing by the median BAL quasar luminosity, $L\approx 3.0\times 10^{46}$ erg~s$^{-1}$ (Section 2.0), then yields a typical minimum ratio of  $L_K/L \gtrsim 0.004$. This lower limit is similar to estimates for individual \pv\ BAL quasars \citep{Capellupo14, Capellupo17, Moravec17} and very near the approximate threshold $L_K/L\gtrsim 0.005$ for quasars playing a major role in disrupting star formation and mass assembly in their host galaxies \citep{Hopkins10}. 

The important new result from our study is that the presence of \pv\ absorption in typical BAL spectra requires large total column densities moving at high speeds. These outflows carry enormous kinetic energies even if they reside at small parsec-scale distances. Larger radial distances, as inferred in some the studies cited above, would increase the estimated kinetic energy luminosities by $L_K\propto R$. However, only a slight increase to $R=1.2$~pc would yield $L_K/L\gtrsim 0.005$, indicating that typical BAL outflows in our study have kinetic energies sufficient for important feedback to host galaxy evolution. If all quasars have BAL outflows \citep[detected via UV absorption lines only in a fraction of quasars due to their small global covering factors,][]{Hamann93}, then this conclusion should be expanded to say that typical \textit{quasars} in our study have sufficient kinetic energy yields for feedback. 

Finally, it is useful to compare the typical mass outflow rate, $\dot{M}_{out}$, to the mass accretion rate, $\dot{M}_{acc}$, using the estimates above. The mass outflow rate is $\dot{M}_{out} = 2L_K/\textrm{v}^2 \gtrsim 5.9$ \msun\ yr$^{-1}$, which is a lower limit based on $L_K\gtrsim 1.2\times 10^{44}$ erg~s$^{-1}$. The mass accretion rate needed to generate the bolometric luminosity is $\dot{M}_{acc} = L/\eta c^2 \approx 5.3$  \msun\ yr$^{-1}$, where we adopt $\eta = 0.1$ as a typical radiative efficiency. Thus we have the interesting result for $\dot{M}_{out} \gtrsim \dot{M}_{acc}$, i.e., the mass outflow rate is at least as large as the mass accretion rate into the black hole. It is important to note that the accretion rate estimated here from the near-UV continuum flux (using a standard quasar SED dominated by emission in the far-UV) pertains to the inner accretion disk where the UV light is emitted. The accretion rates at larger radii, e.g., beyond the regions that emit the UV flux and launch the outflow, must be $\dot{M}_{out} + \dot{M}_{acc} \gtrsim 11.2$ \msun\ yr$^{-1}$ to account for both the outflow and the UV luminosity. 

\subsection{From mini-BALs to BALs to LoBALs}

The persistent presence of \pv\ \lam 1118,1128 absorption in our absorber-frame composite spectra, along with $\sim$1:1 depth ratios in resolved doublets like \pv\ and \siiv, indicate that all of the outflow types from mini-BALs to the strongest BALs have typically large column densities and saturated absorption in most of their UV lines (Sections 5.2 and 5.3). Line saturation, in particular, means that increasingly deeper and broader lines from mini-BALs to strong BALs are caused mainly by larger line-of-sight covering fractions: mini-BALs and weak BALs identify small outflow clumps with low velocity dispersions, while deeper and broader BALs might form in larger clumps or larger numbers of small mini-BAL-like clumps. 

However, the outflows cannot be characterized simply by the sizes and numbers of `clumps' along our lines of sight because the data require inhomogeneous partial covering, with a distribution of column densities across the emission source (Section 5.3). The column densities and covering fractions are therefore closely intertwined. For example, if the line optical depths span a range from thin to thick across the projected area of the emission source, then increasing the column densities and optical depths everywhere by a simple scale factor will cause the lines to become optically thick over larger spatial areas, leading to larger covering fractions and deeper observed line troughs \citep[see also][]{Hamann04, Arav08, Hamann11, Moravec17}. In this way, mini-BALs could become BALs without any changes to the spatial structure (e.g., the number or sizes of `clumps').

Another important result from our study is that the strengths of low-ionisation outflow lines like \aliii\ and \cii\ are closely related to the strength of \pv\ absorption (Figures 6 and 7). For example, the median \pv -selected quasar in our study is a LoBAL based on strong \aliii\ absorption (Section 4.2.3) and, conversely, the median LoBAL quasar has very strong \pv\ absorption (Section 4.2.2). The relationship between \pv\ and the lower ions in the  outflows is also evident from their similar absorber-frame line profiles, which are narrower and less blueshifted than \civ\ in the same spectrum. It is also notable that the LoBAL composite spectra have characteristically strong absorption in high-ionisation lines like \civ , \nv, and \ovi. In fact, the \civ, \nv, and \ovi\ lines are as strong in LoBAL quasars selected to have strong \aliii\ as they are in HiBALs selected to have strong \civ\ (c.f., Figures 4, 5 and 8). 

These results directly connect LoBALs to large outflow column densities and \textit{high} degrees of ionisation. We might naively expect the opposite, that stronger low-ionisation lines in LoBALs compared to HiBALs identify lower ionisation parameters in weaker radiative environments, perhaps farther from the quasars. However, lower ionisation parameters would lead to weaker high-ionization absorption lines like \nv\ and \ovi\ and they \textit{cannot} produce strong \pv\ absorption \citep[e.g.,][Hamann et al., submitted, and refs therein]{Leighly11}. The strong \pv\ lines that accompany strong LoBALs in our composite spectra require \textit{large} ionisation parameters and large total column densities to generate thick layers of fully-ionised gas \citep[see also][]{Baskin14}. This indicates that 1) LoBALs form generally in the same high-intensity radiation environments as HiBALs, and 2) the physical relationship of LoBALs is PV absorption is their shared need for large column densities in harsh radiative environments. Low-ionisation lines like \aliii\ and especially \cii\ cannot form directly in the same gas as \pv\ because they require a less-ionised environment. But they \textit{can} form deep inside high-column density outflow clouds where radiative shielding allows the lower ions to survive behind thick layers of fully-ionised \pv-absorping gas. This stratified ionization structure was described in detail recently by Hamann et al. (submitted, see their Figure 7). We conclude that the stronger low-ionisation lines in LoBALs compared to HiBALs are caused primarily by larger total column densities and/or larger covering fractions in high-column density gas. 

Strong LoBAL quasars also have the weakest \heii\ emission lines and the largest velocity shifts in their \civ\ BALs among the outflow samples in our study (Figure 3, Section 4.1). Orientation effects might play some role in determining the absorption-line characteristics. For example, in unified models of quasar outflows \citep[e.g.,][and refs. therein]{Ganguly99, Elvis00, Elvis12, Hamann12}, LoBALs might appear for sightlines near the accretion-disk plane where there are naturally larger outflow column densities and perhaps more dust reddening associated with a dusty torus \citep[Section 4.1, see also the computational models by, e.g.,][]{Murray95, Murray97, Proga04, Proga12, Matthews16}. However, it is not at all clear that orientation effects can explain the trend for weaker \heii\ \lam 1640 and other high-ionization emission lines in quasars with stronger BALs, culminating in the LoBALs. 

This emission-line behavior, combined with the evidence for larger column densities and larger velocity shifts, suggests that LoBALs identify a more powerful variety of quasar outflow enhanced by softer far-UV spectra and more effective radiative driving (Section 5,1). There is also a weak tendency for LoBALs to have larger near-UV luminosities (Section 4.1.1), which naturally favor higher accretion rates for a given black hole mass (i.e., larger $L/L_E$), faster outflows, and larger mass loss rates \citep[see also][]{Urrutia12, Glikman12, Baskin14}. In our study, the emission-line behavior related to far-UV spectral slope is well-measured and dramatic (Section 4.1.3), while the luminosity trend is weak and the relationship to $L/L_E$ is unknown because we do not measure black hole masses and $L_E$. 

We conclude that LoBALs are at one extreme in a distribution of quasar outflow properties, from mini-BALs to BALs to LoBALs, where more powerful outflows result primarily from softer UV spectral slopes and perhaps, secondarily, from larger luminosities and higher accretion rates across this sequence. 

\section{Summary}

We examine rest-frame UV spectra of 6856 BAL quasars at redshifts $2.3<z<3.5$ selected from the BOSS DR12 quasar catalog. We use median composite spectra in the quasar frame and the absorber frame, covering rest wavelengths from at least 1090 \AA\ to 2250 \AA , to study trends in the data and infer basic physical properties of the outflows. This work extends earlier studies of BAL composite spectra by \cite{Baskin13} and \cite{Baskin15}. We specifically examine quasar subsamples sorted by \civ\ \lam 1549,1551 BAL strength and by LoBAL strength based on \aliii\ \lam 1855,1863 absorption, plus additional subsamples of HiBALs, mini-BALs, and \pv -selected BAL quasars. Our main results are the following:

(1) Stronger/deeper BALs with larger velocity are accompanied by weaker \heii\ \lam 1640 emission-line REWs (Figure 3 and Section 4.1). Consistent with previous studies, we attribute this trend to softer ionising UV spectra that produce weaker \heii\ emission for a given near-UV flux, as well as less ionisation, larger near-UV opacities and more effective radiative acceleration in the outflows (Section 5.1). 

(2) LoBAL quasars have characteristically very strong \civ\ BALs similar to the strongest BALs in our study. They also have the weakest (unmeasurable) \heii\ emission lines, the strongest emission in low-ionisation lines like \feii, unusually strong \pv\ \lam 1118,1128 absorption, the largest velocity shifts in their \civ\ BALs, and the reddest near-UV continua (Sections 4.1 and 4.2.2).  

(3) The flux minima in low-ionisation BALs such as \aliii, \cii\ \lam 1335, and \mgii\ \lam 2798,2803 have smaller blueshifts than \civ\ BALs in the median spectra by typically $\sim$800 to $\sim$2200 \kms\ (Section 4.2.2).  

(4) \pv\  \lam 1118,1128 absorption is present in every composite, from mini-BALs to the strongest BALs, and the doublet has a saturated $\sim$1:1 depths ratio everywhere it is resolved (Figures 5 to 8). This requires \textit{typical} large total column densities $\log N_H (\textrm{cm}^{-2}) \gtrsim 22.7$ (for solar abundances) and large ionisation parameters $\log U \gtrsim -0.5$ across all outflow type (Section 5.2). It also implies that most UV outflow lines are saturated, confirmed by measured $\sim$1:1 depth ratios in other doublets like \siiv\ \lam 1394,1403, \ovi\ \lam 1032,1038, and \aliii\ (Section 5.3). 

(5) Different observed depths in saturated lines identify clumpy inhomogeneous outflow structures that partially cover the near-UV emission source $\lesssim$0.006 pc across. The covering fractions are optical depth-dependent, where weak/low-abundance lines like \pv\  form in small clumps with the highest column densities while strong lines like \civ\ and \ovi\ form also in lower-column density gas that covers both larger spatial areas (to produce deeper troughs) and wider ranges in velocity (to produce broader profiles, Section 5.3). 

(6) The excited-state \siv * \lam 1073 line is stronger than the resonance line \siv\ \lam 1063 in all BAL composites (except the strongest LoBALs where the lines are overwhelmed by blends), indicating that BAL outflows have typical minimum densities $n_e \gtrsim 3\times 10^5$ \cmn\ and maximum  radial distance $R\lesssim 23$ pc from the quasars (Section 5.4). This density result is also supported by detections of highly-excited \ciii * \lam 1076 absorption. 

(7) If the actual radial distances of BAL outflows are $R\sim1.2$~pc, then their typical minimum kinetic energy luminosities relative to bolometric, $L_K/L\gtrsim 0.005\, (R/1.2\textrm{pc})$, are sufficient to drive important feedback effects in the host galaxies (Section 5.5). 

(8) Typical BAL quasars have $\dot{M}_{out} \gtrsim \dot{M}_{acc}$, i.e., mass outflow rates that are similar to or larger than the mass accretion rates into the central black hole (Section 5.5). 

(9) LoBALs are accompanied by stronger \pv\ absorption and all of the usual high-ionisation lines like \civ\ and \ovi\ (Section 4.2.2 and 4.2.3), thus indicating that LoBALs form in harsh radiative environments (with large $U$) like other BALs but where larger total column densities lead to enhanced radiative shielding. Combined with weaker \heii\ emission lines and broader \civ\ BAL troughs (Figure 3, Section 4.1.2 and 4.1.3), LoBALs appear to be at one extreme in a continuum of outflow properties from mini-BALs to BALs to LoBALs where more powerful outflows are produced primarily by softer ionising UV spectra and more effective radiative driving. 

\section*{Acknowledgments}
We are grateful to Alexi Baskin and Ari Laor for valuable discussions on the science and the composite spectrum techniques. We also thank an anonymous referee for helpful comments on the manuscript. FH and HH acknowledge financial support from the USA National Science Foundation via the grant AST-1009628. Funding for SDSS-III has been provided by the Alfred P. Sloan Foundation, the Participating Institutions, the National Science Foundation, and the U.S. Department of Energy Office of Science. The SDSS-III web site is http://www.sdss3.org/. SDSS-III is managed by the Astrophysical Research Consortium for the Participating Institutions of the SDSS-III Collaboration including the University of Arizona, the Brazilian Participation Group, Brookhaven National Laboratory, Carnegie Mellon University, University of Florida, the French Participation Group, the German Participation Group, Harvard University, the Instituto de Astrofisica de Canarias, the Michigan State/Notre Dame/JINA Participation Group, Johns Hopkins University, Lawrence Berkeley National Laboratory, Max Planck Institute for Astrophysics, Max Planck Institute for Extraterrestrial Physics, New Mexico State University, New York University, Ohio State University, Pennsylvania State University, University of Portsmouth, Princeton University, the Spanish Participation Group, University of Tokyo, University of Utah, Vanderbilt University, University of Virginia, University of Washington, and Yale University.

\bibliographystyle{mnras}
\bibliography{../../bibliography}

\bsp

\label{lastpage}

\end{document}